\documentstyle[psfig]{mn} 
\newif\ifAMStwofonts  
  
\newcommand{\Lya}{Ly$\alpha$}  
\newcommand{\OII}{[O{\footnotesize II}]$\lambda$3727}  
\newcommand{\Ha}{H$\alpha$}  
\newcommand{\Hb}{H$\beta$}  
\newcommand{\Hg}{H$\gamma$} 
\newcommand{\Hd}{H$\delta$} 
 \newcommand{\He}{H$\epsilon$} 
   
\newcommand{\etal}{~et~al.~}                                
\newcommand{\kms}{ km\ s$^{-1}$}                            
\newcommand{\ergs}{erg s$^{-1}$}                            
\def\ergseq{{\rm erg}\: {\rm s}^{-1}\:}                 
\newcommand{\Hub}{km\ s$^{-1}$ Mpc$^{-1}$}                  
\newcommand{\Msolar}{M$_{\odot}$}                           
\def\Myeareq{M_{\odot} yr^{-1}}                  

\ifoldfss  
  \ifCUPmtlplainloaded \else  
    \NewTextAlphabet{textbfit} {cmbxti10} {}  
    \NewTextAlphabet{textbfss} {cmssbx10} {}  
    \NewMathAlphabet{mathbfit} {cmbxti10} {} 
    \NewMathAlphabet{mathbfss} {cmssbx10} {} 
  \fi  
  \ifAMStwofonts  
    \ifCUPmtlplainloaded \else  
      \NewSymbolFont{upmath} {eurm10}  
      \NewSymbolFont{AMSa} {msam10}  
      \NewMathSymbol{\upi}     {0}{upmath}{19}  
      \NewMathSymbol{\umu}     {0}{upmath}{16}  
      \NewMathSymbol{\upartial}{0}{upmath}{40}  
      \NewMathSymbol{\leqslant}{3}{AMSa}{36}  
      \NewMathSymbol{\geqslant}{3}{AMSa}{3E}

    \fi  
  \fi  
\fi 
  
\ifnfssone  
  \newmathalphabet{\mathit}  
  \addtoversion{normal}{\mathit}{cmr}{m}{it}  
  \addtoversion{bold}{\mathit}{cmr}{bx}{it}  
  \newmathalphabet{\mathbfit} 
  \addtoversion{normal}{\mathbfit}{cmr}{bx}{it}  
  \addtoversion{bold}{\mathbfit}{cmr}{bx}{it}  
  \newmathalphabet{\mathbfss} 
  \addtoversion{normal}{\mathbfss}{cmss}{bx}{n}  
  \addtoversion{bold}{\mathbfss}{cmss}{bx}{n}  
  \ifAMStwofonts  
    \ifCUPmtlplainloaded \else  
      \UseAMStwoboldmath  
      \makeatletter  
      \new@mathgroup\upmath@group  
      \define@mathgroup\mv@normal\upmath@group{eur}{m}{n}  
      \define@mathgroup\mv@bold\upmath@group{eur}{b}{n}  
      \edef\UPM{\hexnumber\upmath@group}  
      \new@mathgroup\amsa@group  
      \define@mathgroup\mv@normal\amsa@group{msa}{m}{n}  
      \define@mathgroup\mv@bold\amsa@group{msa}{m}{n}  
      \edef\AMSa{\hexnumber\amsa@group}  
      \makeatother  
      \mathchardef\upi="0\UPM19  
      \mathchardef\umu="0\UPM16  
      \mathchardef\upartial="0\UPM40  
      \mathchardef\leqslant="3\AMSa36  
      \mathchardef\geqslant="3\AMSa3E  
    \fi  
  \fi  
\fi 
  
\ifnfsstwo  
  \DeclareMathAlphabet{\mathbfit}{OT1}{cmr}{bx}{it}  
  \SetMathAlphabet\mathbfit{bold}{OT1}{cmr}{bx}{it}  
  \DeclareMathAlphabet{\mathbfss}{OT1}{cmss}{bx}{n}  
  \SetMathAlphabet\mathbfss{bold}{OT1}{cmss}{bx}{n}  
  \ifAMStwofonts  
    \ifCUPmtlplainloaded \else  
      \DeclareSymbolFont{UPM}{U}{eur}{m}{n}  
      \SetSymbolFont{UPM}{bold}{U}{eur}{b}{n}  
      \DeclareSymbolFont{AMSa}{U}{msa}{m}{n}  
      \DeclareMathSymbol{\upi}{0}{UPM}{"19}  
      \DeclareMathSymbol{\umu}{0}{UPM}{"16}  
      \DeclareMathSymbol{\upartial}{0}{UPM}{"40}  
      \DeclareMathSymbol{\leqslant}{3}{AMSa}{"36}  
      \DeclareMathSymbol{\geqslant}{3}{AMSa}{\"3E}  
    \fi  
  \fi  
\fi 
  
\ifCUPmtlplainloaded \else  
  \ifAMStwofonts \else 
    \def\upi{\pi}  
    \def\umu{\mu}  
    \def\upartial{\partial}  
  \fi  
\fi

\title[An Empirical Calibration of Star Formation Rate Estimators]  
{An Empirical Calibration of Star Formation Rate Estimators}  
  
\author[D. Rosa--Gonz\'alez, E. Terlevich and R. Terlevich]  
{Daniel Rosa--Gonz\'alez$^1$,  
Elena Terlevich$^1$\thanks{Visiting Fellow at IoA, UK} and Roberto   
Terlevich,$^{2}$\thanks{Visiting Professor at INAOE, Mexico} \\  
$^1$ INAOE, Luis Enrique Erro 1. Tonantzintla, Puebla 72840. M\'exico.\\   
$^2$ Institute of Astronomy, Madingley Road, CB3 OHA Cambridge, U.K.}  
\date{Accepted  .  
      Received ;  
      in original form \today\~~ SFR-v126}  
  
\pagerange{\pageref{firstpage}--\pageref{lastpage}}  
\pubyear{2001}  
  
\begin{document}  
  
\maketitle  
  
\label{firstpage}  
  
\begin{abstract}  
  
The observational determination of the behaviour of the star formation   
rate (SFR) with look-back time or redshift has two main weaknesses: 
1~- the large uncertainty  
of the dust/extinction corrections, and 2~- that systematic errors may be introduced 
by the fact that the SFR is estimated using different methods at 
different redshifts.  Most frequently, the luminosity of the \Ha\ emission line, that of the   
forbidden line \OII\   
and that of the far ultraviolet continuum (UV) are used with low,   
intermediate and high redshift galaxies respectively. 
 
To  assess  the possible systematic   
differences among the different SFR estimators  and the role of dust, 
we have compared SFR estimates  using \Ha , SFR(\Ha), \OII \AA , SFR(OII), UV, SFR(UV) and  FIR, 
SFR(FIR) luminosities of a sample comprising the 31 nearby star forming   
galaxies having high quality photometric data in the  UV, optical and  FIR.

We review the different ``standard" methods for the estimation of the SFR 
and find that while the standard method 
provides good agreement between SFR(\Ha) and SFR(FIR), both SFR(OII) and SFR(UV) 
are systematically higher than SFR(FIR), irrespective  of the extinction law.  
 
We show that the excess in the SFR(OII) and SFR(UV)  
is mainly due to an overestimate of the extinction resulting from the effect 
of underlying stellar Balmer absorptions in the measured emission line fluxes. 
Taking this effect into consideration 
in the determination of the extinction brings the SFR(OII) and 
SFR(UV) in line with the SFR(FIR) and simultaneously reduces the internal scatter of
the SFR estimations.

Based on these results we have derived ``unbiased" SFR expressions 
for the SFR(UV), SFR(OII) and  SFR(\Ha).
We have used these estimators to recompute  the SFR history 
of the Universe using the results of published surveys. The 
main results are that the use of the unbiased SFR estimators brings into 
agreement the results of all surveys. 
Particularly important is the agreement achieved for the SFR derived from 
the FIR/mm and optical/UV surveys. The ``unbiased"  star formation 
history of the Universe shows a steep rise in the SFR 
from $z=0$ to $z=1$ with SFR $\propto (1+z)^{4.5}$
followed by a decline for $z>2$ where  SFR $\propto (1+z)^{-1.5}$.
Galaxy formation  models tend to have a much flatter slope
from $z=0$ to $z=1$.

\end{abstract}  
  
\section{Introduction}

The knowledge of the history of star formation at cosmic scales is 
fundamental to the understanding of the formation and evolution of galaxies.   
Madau and collaborators combined the results from the ultraviolet surveys 
of Lilly \etal  \shortcite{1996Lilly} with the information from the 
Hubble Deep Field to give an estimate of the star formation 
history from $z = 0$ to $z = 4$. Subsequent studies have indicated 
the crucial role played by dust in the estimates of 
SFR. Large correction factors were suggested for $z > 1$--$2$ 
by several authors (Meurer \etal 1997, Meurer, Heckman and Calzetti 1999, 
Steidel \etal 1999, Dickinson 1998). But even these large dust extinction 
corrections do not seem to be enough to bring the 
optical/UV SFR estimates in line with the mm/sub-mm  ones (Hughes \etal 1998; 
Rowan-Robinson \etal 1997 (RR97); Chapman \etal 2001).

On top of the uncertainties associated with the extinction correction, 
most of  the SFR estimates have been 
performed using expressions   
derived from spectra constructed using population synthesis methods, an  
approach  that requires four rather uncertain ingredients:   
1) an initial mass function (IMF); 
2) a stellar evolutionary model grid giving the luminosity and   
effective temperature as a function of time;   
3) a stellar atmospheres grid that assigns a spectrum to each   
star for a given luminosity and effective temperature, and  
4)  a star formation history. The fact that the redshift evolution   
of the SFR is  constructed using  different estimators at  
different redshift ranges is a potential source of systematic effects with 
redshift that can distort the shape of the evolutionary curve. 
In practice the \Ha\ luminosity is used to estimate the SFR for galaxies  
with redshifts up to 0.4; the [OII]~$\lambda$3727\AA\ line , for those with 
0.4$>$z$>$1.0   
and the UV continuum luminosity for galaxies with z$>$2.0. 
In addition, dust extinction corrections are not 
treated uniformly over the whole redshift range.  
 
It is therefore important to ensure that there are no  
systematic differences between the different estimators and corrections 
that can distort the results. 
 
This paper has two main aspects, in the first 5 sections we  review the  
``standard" methods for the estimation of the SFR  
and test the consistency of the different SFR estimators by applying them to a sample 
of well studied nearby star forming galaxies and comparing  the results. In the absence of 
systematic differences among them, all should give the same SFR for each one of the galaxies in the sample. We then use the results of the nearby sample to construct a set
of ``unbiased'' SFR estimators and apply them to published surveys.

\section{The Reference Galaxy Sample}   
\label{sample}

To critically test the possibility of systematic differences between 
the  SFR estimators we have compiled from the literature a sample of all the   
nearby well studied star forming galaxies for which good 
data is available in \Ha, \Hb, \OII , UV   
continuum and FIR. We call it the   
``reference'' sample of galaxies.  
 
The resulting sample consists of galaxies  
classified either as HII galaxy, Starburst (SB)    
or Blue Compact (BCG) and although it covers a  range of galaxian 
properties, a large fraction of them is 
of low luminosity and low metal content.  
Work by Koo and collaborators (Koo et al.~1996, Lowenthal et al.~1997) 
 has shown 
deep similarities between the faint blue galaxies found typically 
at $z \sim\ 0.5$ and the type of galaxies in our sample. This could also be 
the case for 
Lyman limit galaxies \cite{1996Giavalisco}.

Spectroscopic data for  14 of the galaxies come from  
McQuade, Calzetti and Kinney \shortcite{1995McQuade} where they combined 
spectral   
data in the   
UV from the IUE\ \cite{1993Kinney} with optical observations.   
The optical spectra were obtained with the KPNO 0.9m telescope  
covering a spectral range from 3500 \AA\ to 8000 \AA\ with  10 \AA\   
resolution.   
The circular  aperture used by     
McQuade \etal\ \shortcite{1995McQuade}, 13.5\arcsec \  in diameter, matches 
the  10\arcsec $\times$ 20\arcsec\  IUE aperture. Therefore no area   
renormalization was necessary for these galaxies when comparing UV with optical data.

Optical data for other 12 galaxies were obtained by Storchi-Bergmann, Kinney and Challis 
\shortcite{1995Storchi} and combined with the  \cite{1993Kinney}   
IUE ultraviolet atlas.    
The optical observations were made with the 1m and 1.5m CTIO   
telescopes. They used a long slit of 10 arcsecond width which was   
vignetted to equal the  10\arcsec $\times$ 20\arcsec\ IUE aperture.  
The 1m telescope  covered a range from 3200 to 6400  
\AA\  with a resolution of 5.5 \AA\ and the 1.5m telescope covered   
from 6400 to 10000 \AA\ with a resolution of 8 \AA . In general   
their spectra  flux levels agree within  20\%   
with the spectra from the IUE satellite. In the cases where the   
differences  
were larger (observations made under non-photometric conditions)  
they assumed that the flux given by the IUE observations is   
the correct one.

The optical spectra of CAM0840, TOL1247 and CAM1543   
were observed by Terlevich \etal\ \shortcite{1991Ter}.   
  TOL1247 was observed under photometric conditions with the   
3.6m ESO telescope and an  8$\times$8 arcsec aperture.   
The spectra of CAM0840 and CAM1543 were obtained at Las Campanas   
Observatory with the 2.5m telescope.   
An aperture of 2$\times$4 arcsec and a resolution of about  
5 \AA\ were used. The ultraviolet spectra of these galaxies   
are from IUE  satellite observations, Terlevich \etal\   
\shortcite{1993Terlevich} (CAM0840 and  TOL1247)   
and Meier \& Terlevich \shortcite{1981Meier} (CAM1543).

The optical data of MRK309 is from the UCM objective-prism   
survey \cite{1996GallegoCAT}.   
The UV data comes  
from the observations made with IUE and reported by McQuade \etal  
\shortcite{1995McQuade}.

Table~\ref{DATA} shows the complete sample of star forming galaxies used in this study. 
All the objects are located at large galactic latitudes ($|$b$|>$25). 
The apparent blue magnitudes are from De Vaucouleurs et al.~\shortcite{RC3}  and 
are corrected for  galactic extinction, internal extinction   
and aperture differences. Radial velocities were corrected by   
the Local Group motion using the NED\protect\footnote{NASA/IPAC Extragalactic   
Database} velocity correction software (Table \ref{DATA}).

The 60 $\mu$m IRAS data for all the galaxies come from   
Moshir \etal\shortcite{IRAS}.

We computed the equivalent width of \Hb\ using the easily deduced expression,

\begin{equation}\large  
\begin{array}{ll}

\label{EW}  
EW(H\beta) = {L(H\beta) \over L_{c}(4861 \AA)} \\ 
~~~~~~~~~~~~~=  2.5 \times 10^{-32}~ \frac{L(H\beta)}{{\rm erg\ s}^{-1}}~  10^{(0.4 M_B)}

\end{array}  
\end{equation}  

\noindent where $L(H\beta)$ and $L_{c}(4861 \AA)$ are the \Hb\ and adjacent continuum 
luminosities respectively.
The field of view used to estimate $M_B$ is larger than the apertures  
for the spectroscopic observations, therefore the estimated EW(\Hb)   
represents a lower limit. Where $M_B$ was not available, as for   
CAM0840, CAM1543 and TOL1247, we used EW(\Hb) from the spectrophotometry of 
Terlevich  \etal\  (1991).


The extinction correction to the observed fluxes both on the continuum 
and on the emission lines was estimated following standard procedures. 
Two different extinction curves were used: the Milky Way   
extinction law  (MW) given by Seaton \shortcite{1979Seaton}  
and Howarth \shortcite{1983Howarth} and the Large Magellanic   
Cloud one (LMC)  
given by Howarth \shortcite{1983Howarth}. A detailed description 
of the procedure can be found in Appendix 1.

 
\begin{table*}   
\begin{center}   
\begin{tabular}{lcrrrrrrrrrr}\hline  
\makebox[0.3cm]{}  & &    & Velocity  & F(\Ha)   & F(\Hb)  &F(\Hg)   & F(OII) & F$_\lambda$(UV)  &  F$_\nu$   &\  \\    
 Name & Type & m$_B$ &  (\kms ) &   &   &  &   & ($\ast$)  &(60 $\mu$m)   & Ref.  \\ \hline   
NGC7673 & HII & 12.86 & 3673.20 & 608.0 & 109.8 &22.1 & 446.6 & 14.6 & 4.95 &      (a)  \\  
CAM0840 & HII & --   & 9000.00 & 127.1 & 37.1 &15.8 & 46.1 & 3.6     & 0.29 &      (b) \\  
CAM1543 & HII & --   & 11392.11 & 191.0 & 52.5 & 22.6 &31.3 & 4.1    & --   &      (e)  \\  
TOL1247 & HII & --   & 14400.00 & 504.5 & 134.9 &58.7 & 159.2 & 11.0 & 0.51 &      (b)\\  
NGC1313 & HII & 9.29 & 269.86 & 148.4 & 17.3 &3.5 & 105.4 & 6.60 & 14.56    &      (d)  \\  
NGC1800 & HII & 12.87 & 614.07 & 125.9 & 26.0 &4.0 & 199.3 & 15.00 & 0.77   &      (d) \\  
ESO572 & HII & 14.16 & 871.51 & 647.9 & 106.7 &52.8 & 283.7 & 19.95 & 0.86  &      (d) \\  
NGC7793 & HII & 9.37 & 253.35 & 221.9 & 33.0 & --  &110.6 & 9.37 & 8.89     &      (d)  \\   
UGCA410 & BCDG & 15.45 & 854.20 & 324.4 & 80.2 &21.5 & 104.6 & 10.4 & 0.30  &      (a) \\  
UGC9560 & BCDG & 14.81 & 1305.00 & 529.4 & 144.6 &50.7 & 334.5 & 18.4 & 0.71 &     (a)  \\  
NGC1510 & BCDG & 13.45 & 737.95 & 512.1 & 116.8 &31.6 & 340.9 & 16.52 & 0.89 &     (d)   \\  
NGC1705 & BCDG & 12.58 & 401.23 & 434.1 & 130.0 &22.3 & 265.0 & 93.67 & 0.87 &     (d)   \\  
NGC4194 & BCG & 12.86 & 2598.46 & 1946.6 & 239.1 &70.1 & 385.3 & 14.2 & 23.52 &    (a)   \\  
IC1586 & BCG & 14.74 & 6045.25 & 229.3 & 38.8 &8.6 & 138.8 & 5.0     & 0.96  &     (a)   \\  
MRK66 & BCG & 15.00 & 6656.66 & 121.3 & 43.2 &11.0 & 148.2 & 9.9      & 0.54 &     (a)  \\  
Haro15 & BCG & --   & 6498.24 & 301.1 & 81.0 &25.7 & 264.6 & 18.15    & 1.35 &     (d)  \\  
NGC1140 & BCG & 12.56 & 1503.13 & 1400.0 & 350.3 &121.3 & 1010.0 & 44.15 & 3.36 &  (d)   \\  
NGC5253 & BCG & 10.87 &270.61 & 7717.0 & 2406.0 &973.2  &4370.3 & 99.05 & 30.51 &  (d) \\  
MRK542  & BCG & 15.80 &7518.56 & 88.9 & 17.1 & --  &3.8  & 5.75 & 0.48 &           (a)\\   
NGC6217 & SB & 11.66 & 1599.81 & 607.4 & 92.9 & 14.8 &108.1 & 15.3 & 11.05 &       (a) \\  
NGC7714 & SB & 12.62 & 2993.82 & 2795.8 & 539.9 &196.9  &951.1 & 26.5 & 10.44 &    (a)  \\  
NGC1614 & SB & 13.28 & 4688.15 & 1069.4 & 92.0 &19.8  &95.2 & 4.8 & 32.71 &        (a)  \\  
NGC6052 & SB & 13.40 & 4818.45 & 565.0 & 122.7 & 39.6 &376.2 & 9.9 & 6.31 &        (a)  \\  
NGC5860 & SB & 14.21 & 5532.07 & 296.7 & 26.8 &-- & 15.4 & 5.5 & 1.64 &            (a) \\  
NGC6090 & SB & 14.51 & 8986.89 & 675.3 & 123.7 &42.0 & 153.6 & 9.6 & 6.45&         (a)   \\  
IC214 & SB & 14.16 & 9161.10 & 152.2 & 21.4 & 5.4& 35.9  & 6.2 & 5.22   &          (a) \\  
MRK309 & SB & 14.61 & 12918.07 & 108.0 & 16.2 &-- &4.4  & 2.4 & 3.43     &         (c)  \\  
NGC3049 & SB & 12.77 & 1321.32 & 513.1 & 116.1 & 51.9 &148.5  & 10.25 & 2.82 &     (d)  \\  
NGC4385 & SB & 12.90 & 1981.44 & 950.0 & 150.7 &51.2 & 261.0 & 11.82 & 4.73 &      (d)  \\  
NGC5236 & SB & 7.98 & 304.31 & 4507.0 & 940.1 & 154.7 & 440.0  & 185.47 & 110.30 & (d)  \\  
NGC7552 & SB & 11.13 & 1571.17 & 2064.0 & 277.8 & 44.5& 243.8  &  19.97 & 72.03 &  (d)  \\ \hline  
\end{tabular}   
\caption{\label{DATA}  
Observed properties of the galaxy sample.  
The blue apparent magnitudes are from De Vaucouleurs et al. 
\protect\shortcite{RC3}.     
The emission lines and the ultraviolet continuum fluxes   
are from    
(a)  McQuade\etal\protect\shortcite{1995McQuade},  
(b)  Terlevich\etal \protect\shortcite{1993Terlevich},  
(c)  Gallego\etal\protect\shortcite{1996GallegoCAT} and   
     McQuade\etal\protect\shortcite{1995McQuade},   
(d)  Storchi-Bergmann\etal\protect\shortcite{1995Storchi},  
(e)  Meier $\&$ Terlevich \protect\shortcite{1981Meier} and   
     Terlevich\etal\protect\shortcite{1991Ter}.   
Units for the intensities are 10$^{-15}$   
erg s$^{-1}$ cm$^{-2}$ except  ($\ast$) where  the  units are 10$^{-15}$    
erg s$^{-1}$ cm$^{-2}$\AA$^{-1}$. The FIR fluxes are IRAS at 60 $\mu$m 
 in Janskys.}   
\end{center}   
\end{table*}   
 

\section{Standard SFR Estimates}   
\label{methods}   
   
In this section we describe the four estimators we have used to 
compute the SFR from the luminosities of the \Ha\ and \OII\ nebular lines,  
the UV luminosity and the FIR continuum.   
The expressions used here are the ones we found most frequently referred  
to in the literature. 
All expressions are for  a Salpeter IMF (N(m)$\propto m^{-2.35}$) with    
masses varying from 0.1 to 100 \Msolar, solar metallicity   
and continuous star formation.  
To convert the fluxes into   
luminosities we used a Hubble constant of 70 \Hub.

\subsection{From \Ha\ Luminosities.}

For the estimate of the SFR from \Ha\ luminosities we used the expression   
given by Kennicutt, Tamblyn 
and Congdon \shortcite{1994Kennicutt},   
\begin{equation}   
SFR(H\alpha )({\Myeareq})= 7.9\times  
10^{-42}L_{H\alpha}{(\ergseq)}   
\label{eq_Ha}   
\end{equation}   
  
\noindent valid  for a T$_e$=10$^4$K and  Case B recombination, i.e. all the    
ionizing photons are processed by the gas.

\subsection{From  \OII\ Luminosities.}      
  
The doublet \OII\  luminosity is used as a SFR tracer for objects  
with redshift  
larger than 0.4 where  \Ha\ is shifted outside the optical range.    
Unlike  \Ha , the \OII\  intensity depends not  
only on the electron temperature and density  but also  
on the degree of ionization and on the metallicity of the gas.   
In practice a semi-empirical approach is used combining the 
SFR(\Ha) with the average \Ha\ / \OII\ ratio  given by  
Gallagher\etal\shortcite{1989Gallagher} using a sample of    
75 blue galaxies and by Kennicutt \shortcite{1992Kennicutt}   
from  a  sample of 90 normal and irregular galaxies.

\begin{equation}   
\label{eq_OII}   
SFR([O{\mbox{\footnotesize II}}] )({\Myeareq })= 1.4\times 10^{-41}L_{OII}{(\ergseq)} 
\end{equation}

\subsection{From UV Continuum  Luminosities.}      
  
The UV continuum luminosity is used as a SFR tracer in objects with redshift  
higher than 1--2. At these redshifts all strong emission lines, apart  
from \Lya , are shifted outside the optical range.  
 
In young stellar clusters, the UV spectrum is  dominated by the continuum emission of    
massive stars.    
In evolutionary synthesis models of starbursts, after a short initial transient phase, 
the UV luminosity per unit frequency becomes proportional to the SFR,   
 
\begin{equation}   
\label{eq_UV}   
SFR(UV)({\Myeareq }) = 1.4 \times 10^{-28}L_\nu({\ergseq  
Hz}^{-1})   
\end{equation}

This equation is valid from 1500 to 2800 \AA\ where the integrated spectrum is  
nearly flat in F$\nu$ for a Salpeter IMF, continuous mode of star formation and solar metallicity.   
This region of the UV is not affected by the    
\Lya\ forest and the contribution from old populations    
is still very small (Kennicutt 1998, Madau et al. 1998). 
 
This SFR indicator though, is extremely 
sensitive to uncertainties in the reddening correction.

\subsection{From FIR.}  
 
To transform the observed $60\mu$m luminosity to SFR 
we assumed that a fraction $\epsilon$  of the ultraviolet/optical 
flux emitted by stars is absorbed by 
dust and reemitted  as thermal emission in the far infrared   
(10 -- 300 $\mu$m). Work by, among others, Mas-Hesse and Kunth 
\shortcite{1991Mas-Hesse} 
indicate that even for a small amount of reddening the 
fraction  $\epsilon$ is very close to unity. This, plus the fact 
that no dust extinction correction is necessary, justifies the 
assumption that the FIR luminosity is an excellent indicator 
of the total UV/optical emissivity of a galaxy. 
Thus, the relation between the total FIR luminosity 
and the star formation rate (Kennicutt 1998) is   
  
\begin{equation}  
\label{SFRIR}  
SFR(FIR)({\Myeareq}) = 1/ \epsilon\  4.5 \times 10 ^{-44}   
L_{FIR}(\ergseq)  
\end{equation}  
 
\noindent  
where the  FIR luminosity is given by  L$_{FIR}\sim 1.7\times L_{60\mu m}$  
\cite{2000Chapman}.

A difficult parameter to quantify is the fraction of ionizing photons 
that escape from the nebula. 
Heckman et al (2001) show that in five of the UV brightest local starburst 
galaxies the fraction of photons escaping is less than about 6\%  
while Steidel, Pettini and Adelberger (2001) claim a higher 
escape fraction. 
Bearing in mind that the two samples differ, these results may 
not be contradictory. Tenorio-Tagle et al.  (1999) have shown that 
the escape of photons 
from a starburst may be time dependent with a very large escape probability 
during the most luminous phases and little escape at other stages.

It is important to point out that  if the dust and ionized gas distributions are similar, 
i.e. they coexist spatially, the FIR and the emission line luminosities
will be similarly affected by the escape of photons. 
Thus, under this condition, the ratios of emission line fluxes to FIR flux  
are, to first order, independent of the fraction of escaped photons 
and therefore not very sensitive to variations in the photon escape 
from nebula to nebula.

\section{Comparing the different SFR estimators}  
\label{Discussion}  
 
We have applied the commonly used SFR estimators  
to our reference sample of star forming galaxies. 
Given that the estimators are all for the same IMF and stellar models 
we do not expect  these aspects to introduce any scatter. 
Figure~\ref{SFRvsSFR} shows the SFR(\Ha), SFR(OII) and SFR(UV) plotted against
the SFR(FIR). Clearly the sample shows a correlation plus a large scatter.

To simplify the analysis and simultaneously make use of the fact that SFR(FIR) 
is probably the best SFR estimator available, we will use in  what follows 
FIR normalized SFR, i.e.~the SFR relative to SFR(FIR). 
The  FIR normalized {SFR(H$\alpha$),  SFR(OII) and SFR(UV) are:

\vskip .3cm 
$\Delta H\alpha$~ = log~$\frac{SFR(H\alpha)}{SFR(FIR)}$ 
\vskip .3cm 
$\Delta OII$ = log $\frac{SFR(OII)}{SFR(FIR)}$ 
\vskip .3cm 
$\Delta UV$ = log~$\frac{SFR(UV)}{SFR(FIR)}$ 
\vskip .5cm 

This is better seen in the distribution histograms of the normalized SFR   
as shown in Figure~\ref{F:NoCor}.  
As reference we included in parts a,d,g of the figure 
the normalized SFR computed using the observed luminosities. 
The central and right columns show the dust extinction 
corrected ratios 
using the MW  and the Calzetti extinction laws respectively. 
The corrections were applied following the common methodology and are described in 
Appendix  \ref{SEC:CORR}. 
 
Our main conclusion is that irrespective of the extinction law applied, 
the SFR(\Ha) is close to the SFR(FIR) while both SFR(OII) and SFR(UV) show a clear 
excess. The excess is much larger for SFR(UV) than for SFR(OII)  suggesting 
a wavelength dependent effect, probably an extinction over-correction. 
Bearing in mind that our reference sample has a large fraction of low metallicity 
and low extinction galaxies this result suggests that applying these 
standard methods to estimate 
SFR will systematically overestimate the SFR in samples at intermediate 
and high redshifts where either SFR(OII) or SFR(UV) are used. 
This result is in apparent contradiction with what has been found  and shown in 
Madau-type plots in recent years, where the SFR obtained from UV and optical data are much lower 
than that obtained 
from mm and sub-mm observations at intermediate and high redshifts. In order to reach agreement 
between both determinations, fixed (and somehow arbitrary) amounts of extinction have been 
applied to the UV/optical data, because at the moment, 
the intermediate 
and high redshift samples do not allow a reliable  determination of 
the dust extinction. 
It is worth noting  that Steidel  \etal (1999) applied a fixed correction 
to their sample that is close to the average $\Delta UV$. 
On a positive note we should indicate that the application of the reddening 
corrections reduce considerably the scatter in all three normalized SFR 
estimators as we will show below. 
 
It is important to clarify the origin of the detected excess in the extinction 
corrected $\Delta$OII and $\Delta$UV. 
There is one effect that is not normally taken 
into account, namely that the presence of an underlying young 
stellar population with deep Balmer absorptions will bias the observed 
emission line ratios towards larger Balmer decrement values, mimicking the dust extinction 
effect. In the next section  we will recalculate the SFR for the 
different tracers but including an 
estimation of the effect of an underlying population. We will also estimate the effect 
of photon escape in $\Delta$UV.

\begin{figure*} 
\setlength{\unitlength}{1cm}    
\vspace{8. cm} 
 
\includegraphics{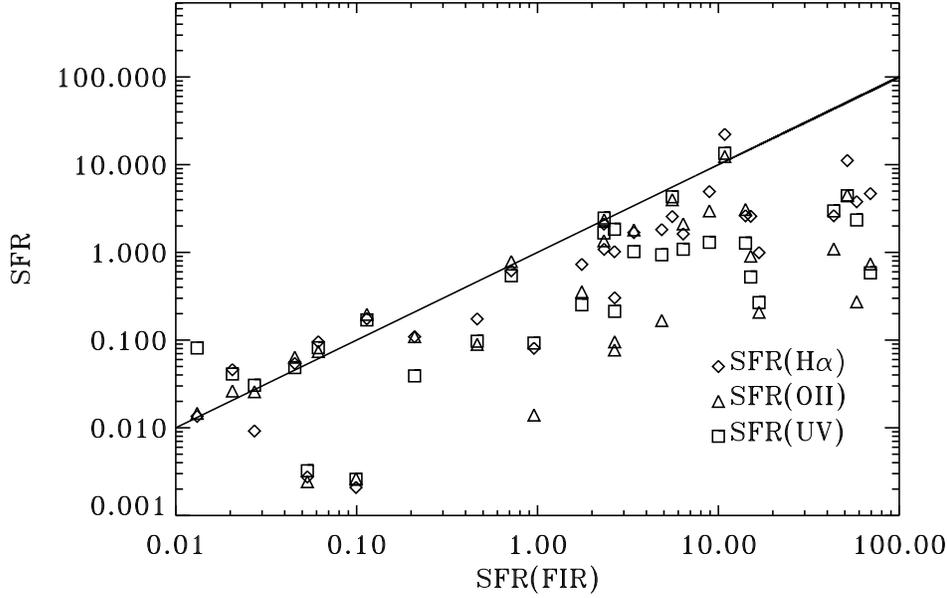}

\caption{\label{SFRvsSFR} Standard SFR estimators vs. SFR(FIR). No extinction corrections
were applied to the data.  The solid line represents equal values.} 
\end{figure*} 
 

\begin{figure*} 
\setlength{\unitlength}{1cm}    
\vspace{10.5 cm} 
\includegraphics{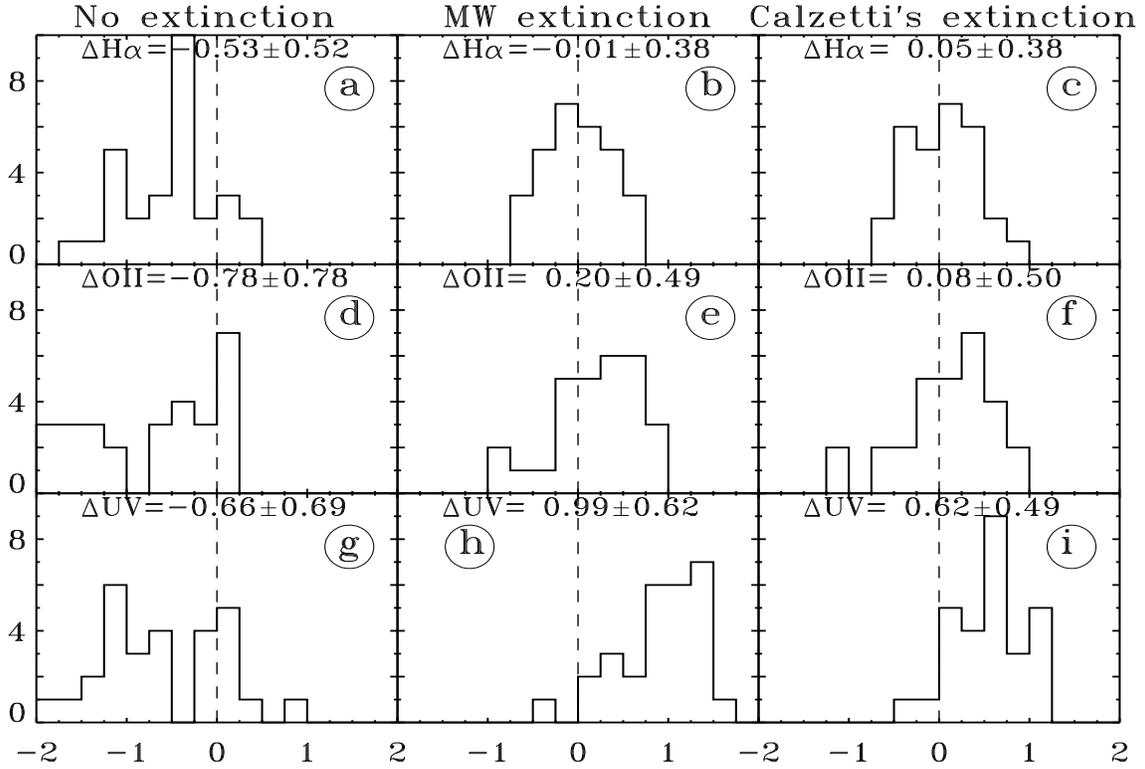} 
\caption{\label{F:NoCor} Histograms of  
the SFR rates given by the different tracers normalized to   
the SFR(FIR).  
In the left panels (a,d,g) no corrections were applied to the data.   
In the central  panels (b,e,h) we corrected the \Ha,  \OII\ and UV continuum 
by using the MW extinction curve.  
In the right panels (c,f,i) the \Ha,  \OII\ and UV continuum were corrected 
with Calzetti's extinction  law. 
The median  
and standard deviation are given for each case. The number of objects is 29.}  
\end{figure*} 
 
\section{Underlying population and Photon Escape} 
 
\subsection{Underlying absorption corrections to the emission line fluxes}  
\label{under}

A clear signature of a population of  
young and intermediate age stars is the presence of the  Balmer 
series in absorption in their optical spectrum. 
A complication is that in star forming  objects, 
the Balmer emission lines from the ionized 
gas appear superimposed 
to the stellar  absorption lines. 
This effect, growing in importance towards the higher order Balmer lines, 
is illustrated in Figure~\ref{F:n1510} where an example (NGC~1510) is
shown. It can be seen that, while \Hb\ emission is moderately affected 
by the absorption, all of the \Hd\ emission is lost into the absorption. 
The equivalent width  of the Balmer absorptions peaks at \Hd\ --  \He\ and 
there is no detection of any absorption in 
\Ha . This is due to the facts that the 
\Ha\ absorption equivalent width is much smaller than that 
of \Hb , and that the wings of the \Ha\ absorption are 
difficult to detect due to the presence of forbidden [NII] doublet 
emission at $\lambda$ 6548\AA\ and $\lambda$ 6584\AA , right on top of both wings. 
In spectra of poorer S/N or lower spectral resolution 
than that of Figure~\ref{F:n1510},  the wings of the Balmer 
absorptions are not detected and the 
result is an underestimate of the emitted fluxes and, 
more important for luminosity determinations, an  
overestimate of the internal extinction (Olofsson 1995).

\begin{figure*} 
\setlength{\unitlength}{1cm}    
\vspace{11.0 cm} 
\includegraphics{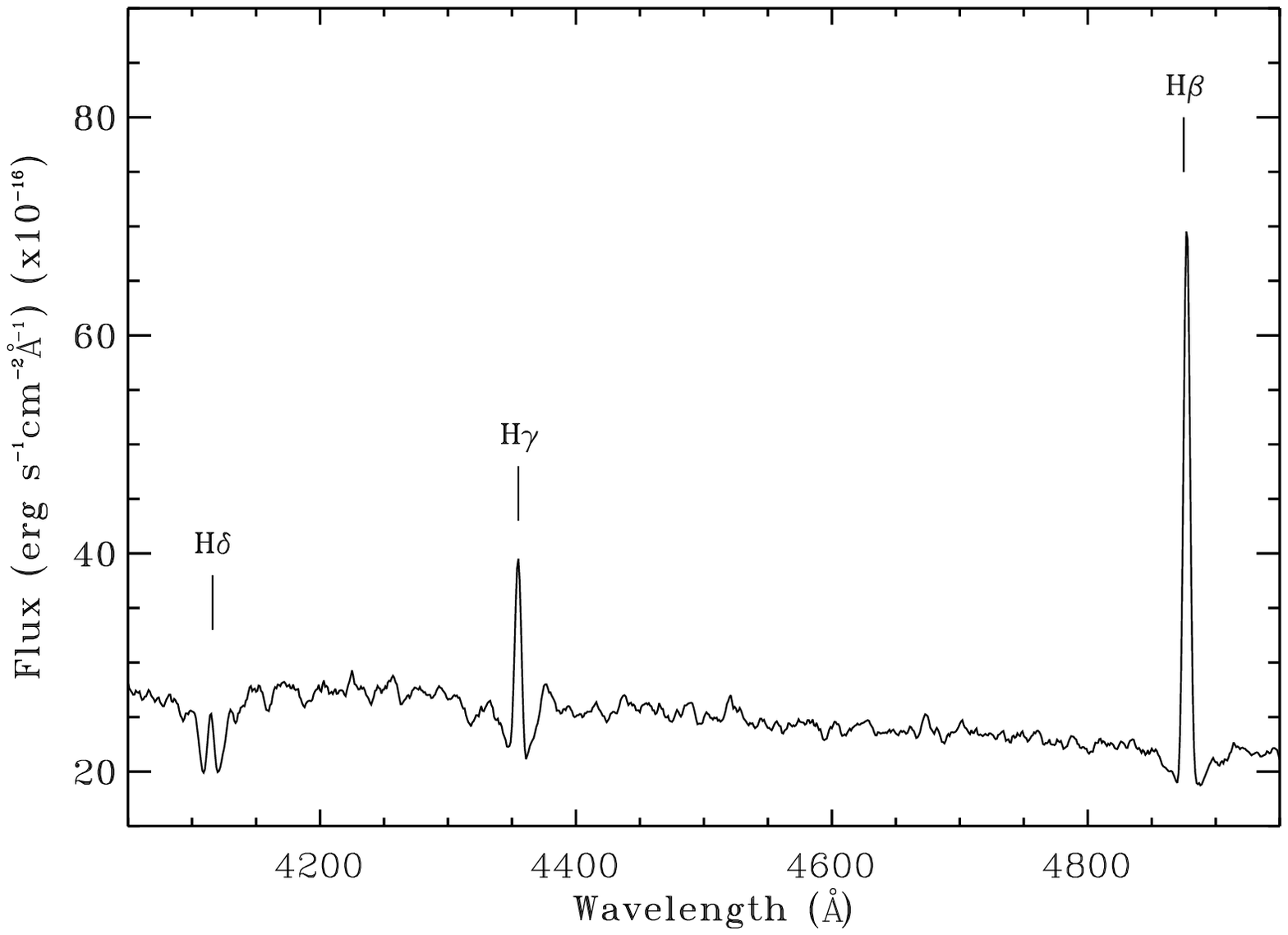} 
\caption{\label{F:n1510} 
A blue spectrum of the star forming galaxy NGC~1510 is shown to illustrate 
the effect of stellar Balmer absorptions in the measurement of 
the emission line strengths. 
} 
\end{figure*} 

\begin{figure*} 
\setlength{\unitlength}{1cm}    
\vspace{8. cm} 
 
\includegraphics{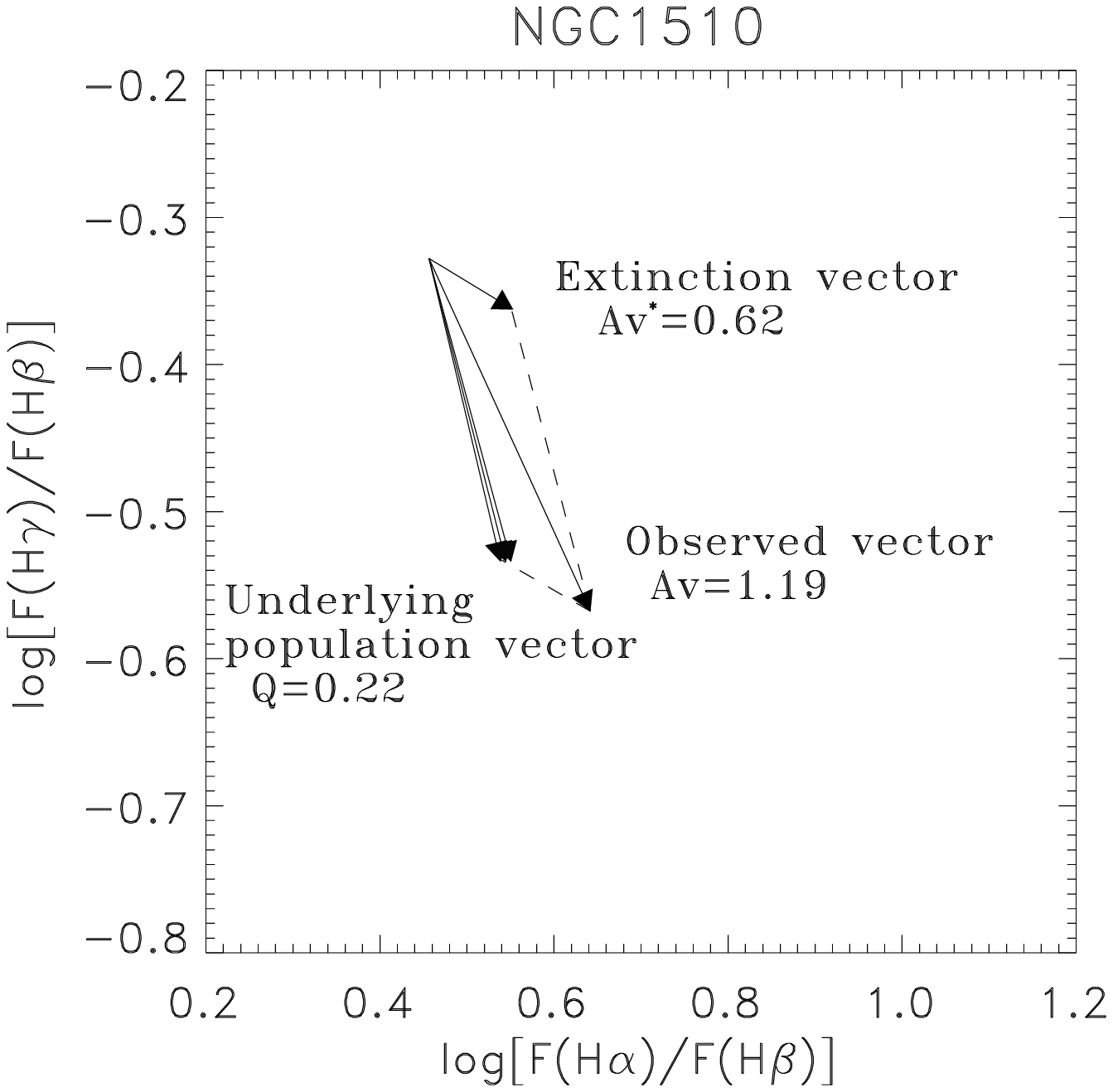} 
\includegraphics{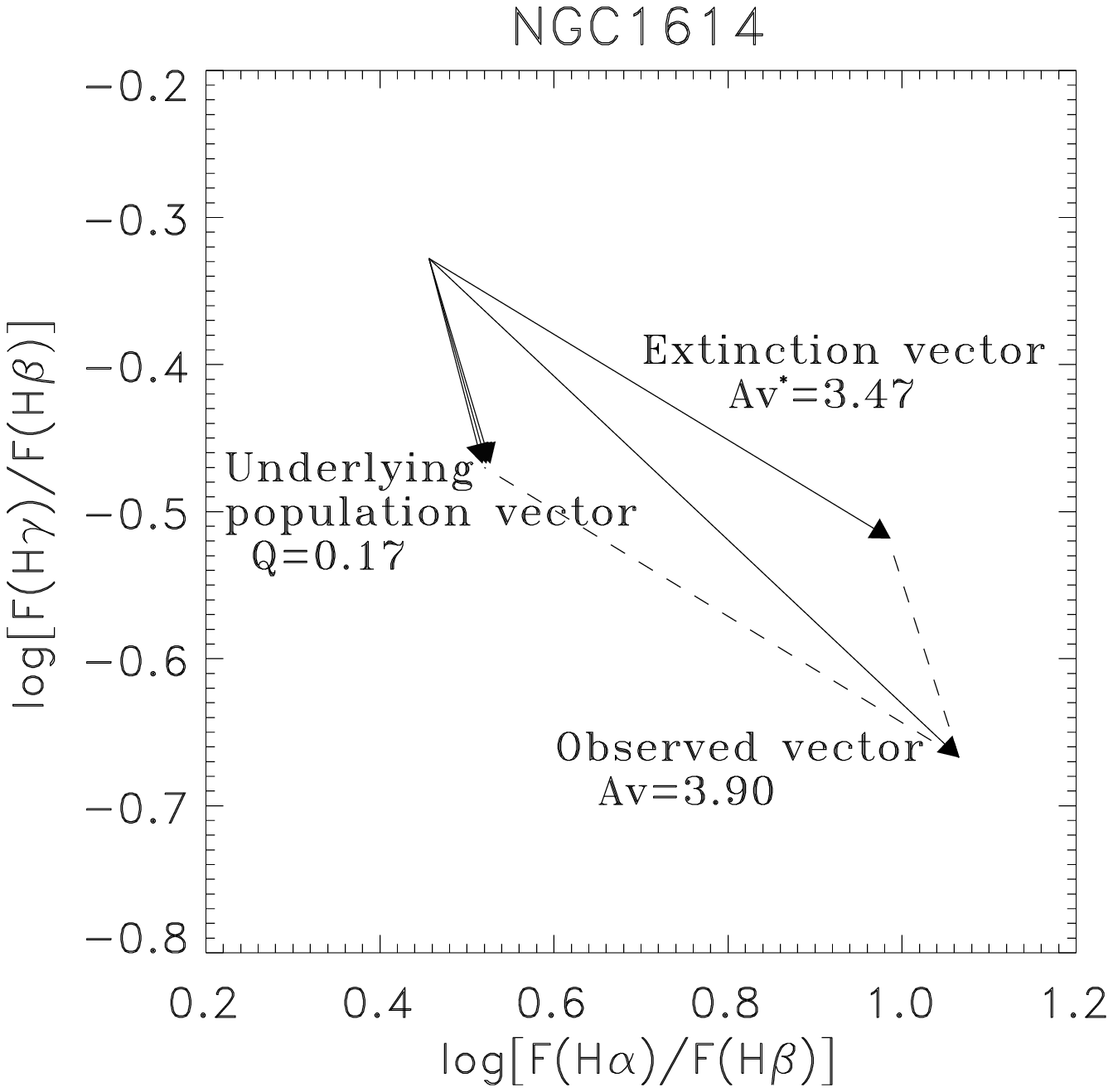} 
 
\caption{\label{both} Logarithmic ratio of F(\Ha)/F(\Hb) vs.  F(\Hg)/F(\Hb).  
The ``observed vector'' (from the intrinsic  
values given by recombination theory to the observed ratio)  
can be decomposed  
in two vectors, one is due to pure extinction and is given by  Equation   
\ref{VEC_EQ}, the other one is given by Equations \ref{HaHbII}   
and \ref{HgHbII} and shows the effect of an underlying   
stellar population which is characterized  by  $Q$ (see text).  
Plotted are the cases for  NGC 1614 and NGC 1510. 
NGC 1614 is an example where the underlying absorption 
correction to the extinction estimate is not very large. NGC 1510 
on the other hand, has a large correction.} 
\end{figure*} 
 
 
The observed ratio  
between two  emission lines (e.g.~\Ha\ and \Hb ), 
when the underlying  absorption is included, is: 
 
\begin{equation}  
\label{Ha}  
\frac{F(H\alpha)}{F(H\beta)}=\frac{F_+(H\alpha) - F_-(H\alpha)}  
{F_+(H\beta) - F_-(H\beta) }  
\end{equation}  
where $F_+(H\alpha)$ and $F_+(H\beta)$ are the intrinsic emission line  
fluxes and $F_-(H\alpha)$ and $F_-(H\beta)$ are the intrinsic fluxes of  
the corresponding absorption lines.  
This expression is correct in the case that the emission and the   
absorption lines have approximately equal widths.  
  
Including the relation between the equivalent width, the flux of  
the continuum and the intensity of the line in equation \ref{Ha}   
we obtain,

\begin{equation}  
\label{HaHbII}  
\frac{F(H\alpha)}{F(H\beta)}=\frac{2.86 -  PQ   
\frac{F_C(H\alpha)}{F_C(H\beta)}}{1- Q}  
= \frac{2.86 \left[1-PQ\frac{EW_+(H\beta)}{EW_+(H\alpha)}\right]}{1-Q}  
\end{equation}  
where,  
$F_C(H\alpha) $ and $F_C(H\beta) $ are the continuum in \Ha\ and \Hb\  
respectively, EW$_+$ and  EW$_-$ are the  
equivalent widths in emission and in absorption respectively for the different lines,  
$Q=\frac{EW_-(H\beta)}{EW_+(H\beta)}$ is the ratio between   
the equivalent widths of \Hb\ in absorption and  in emission, 
$ P= \frac{EW_-(H\alpha)}{EW_-(H\beta)}$ is the ratio between the equivalent   
widths in absorption of \Ha\ and \Hb\ and $F_+$(\Ha)/ $F_+$(\Hb)=2.86 is the 
theoretical ratio between \Ha\ and \Hb\ in emission for Case B recombination  \cite{Osterbrock}.

The value of {\it P} can be obtained from  spectral evolutionary   
calculations like those of Olofsson~\shortcite{1995Olofsson}.  
For the case of solar abundance   
and stellar  masses   
varying between 0.1 and 100~\Msolar\ within a Salpeter IMF,    
the value of {\it P} changes between 0.7 and 1 for ages between    
1 and 15 million years respectively.  
This variation in the {\it P} parameter produces a   
change in the estimated $F$(\Ha)/$F$(\Hb) ratio of less   
than 2\%, so in what follows we asssume    
{\it P}$=$1.  
  
For an instantaneous burst,   
the ratio $\frac{EW_+(H\beta)}{EW_+(H\alpha)}$  varies  between   
0.14 and  0.26 (Mayya \shortcite{1995Mayya}, Leitherer \& Heckman  
\shortcite{1995Leitherer}).  
  
The corresponding equation for  \Hg\ and \Hb\ is:  
  
\begin{equation}  
\label{HgHbII}  
\frac{F(H\gamma)}{F(H\beta)}=\frac{0.47 -GQ}  
{1 - Q}  
\end{equation}  
where $G=\frac{EW_-(H\gamma)}{EW_-(H\beta)}$ is the ratio between the    
equivalent  
widths in absorption of  \Hg\ and \Hb\ and we assume for the respective emissions an intrinsic ratio  
of 0.47 for Case B recombination \cite{Osterbrock}.   
  
The evolution of the   equivalent width  of the Balmer absorption  lines has 
been  analyzed by Gonz\'alez Delgado, Leitherer and Heckman 
\shortcite{1999Gonzalez}.  
In their models  the parameter $G$ is almost constant in    
time and  independent of the adopted  star formation history.   
We fixed the  value of $G$ to 1 as suggested by their results.

The effect of the underlying stellar absorptions is shown as a vector $Q$ 
in Figure~\ref{both} (from equations  \ref{HaHbII} and  \ref{HgHbII}). 
The whole time dependence is shown by the three closely grouped vectors. 
Its range is much smaller than typical observational errors. 
Dust extinction is also represented by a vector 
in the same plane (equation \ref{VEC_EQ}). 
It is possible, as  these two vectors are not parallel, to solve 
simultaneously for underlying absorption ($Q$) and extinction  (Av)  for every 
object for which ${F(H\gamma)}$, ${F(H\beta)}$ and  ${F(H\alpha)}$ are measured.

We further illustrate the presence of underlying Balmer absorption in 
star forming galaxies in Figure~\ref{Scatter} where we have plotted 
the galaxies from our sample in the 
log $\big($F(\Hg)/F(\Hb)$\big)$ \ vs. log $\big($F(\Ha)/F(\Hb)$\big)$ plane. 
Also shown are the vectors depicting  dust extinction and the underlying absorption. 
Clearly, most observational points 
occupy the region below the reddening vector and to the right of the 
Balmer absorption vector. In the absence of underlying absorption 
all points should be distributed along the extinction vector. The 
fact that there is a clear spread below the extinction vector gives 
support to the underlying absorption scenario. 
We also find that the objects 
with smaller equivalent width of \Hb\ are systematically 
further away from the pure extinction vector. 

Four galaxies (NGC~3049, ESO~572, MRK~66 and NGC~1705) fall outside the 
space defined by the extinction and underlying absorption vectors, although 
two of them are within the errors. The other two (ESO~572 and MRK~66) are 
faint and reported as having been observed in less than optimum conditions 
in the original observations paper.

\begin{figure*} 
\setlength{\unitlength}{1cm}    
\vspace{8 cm} 
\includegraphics{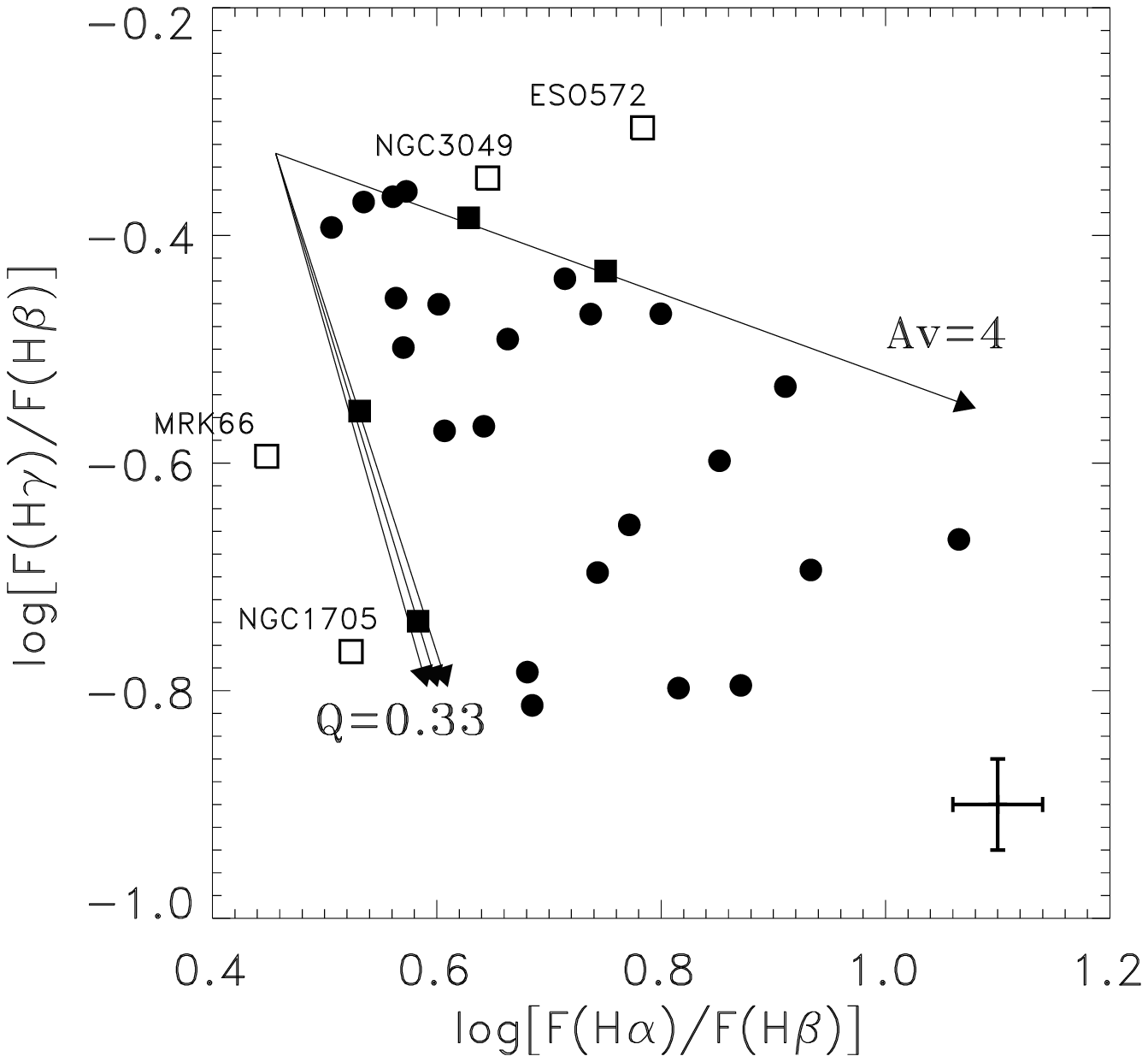} 
\caption{\label{Scatter} The galaxies from our sample are plotted in the 
Balmer decrement plane  log (F(\Ha)/F(\Hb)) vs. log (F(\Hg)/F(\Hb)).  
The vectors indicate the direction of shifts due to extinction or underlying absorption 
from the intrinsic  values given by recombination. The cross indicates typical 
1$\sigma$ errors in the line ratios.  Most points are, within the errors, below the 
reddening line and to the right of the underlying Balmer absorption line.} 
\end{figure*} 
 

We have used  this method to estimate simultaneously the   
``real'' visual extinction Av$^\ast$ and the underlying Balmer absorption $Q$ 
\footnote{See Appendix \ref{SEC:CORR} for a detailed discussion on the dust 
extinction corrections to the observed fluxes.}.  
The values of Av$^\ast$ were then applied to  the UV continuum and the 
emission line fluxes; the corrected values are listed in Table \ref{AvAv}.

\begin{table}  
\begin{center}  
\begin{tabular}{lcccccc}\hline  
 Name     & Av    & Av$^\ast$ & $Q\times$ 100  &   $\beta$   & $(1-\epsilon)$ & EW(H$\beta$)  \\\hline  
  NGC7673 &  1.84 &  1.11     & 27         &   -1.50 &  0.11    &      4.69 \\ 
  CAM0840 &  0.50 &  0.43     &  3       &      -1.26$\dag$ &  0.33    &       121$\star$\\ 
  CAM1543 &  0.67 &  0.67     &  0       &      -0.70$\dag$ &  --     &     224$\star$\\ 
  TOL1247 &  0.75 &  0.75     &  0         &    -0.47$\dag$ &  0.46    &     97$\star$\\ 
  NGC1313 &  3.06 &  2.45     &  23       &     -0.60$\ddag$ &  0.02    &     0.03\\ 
  NGC1800 &  1.47 &  0.56     &  33      &      -1.65 &  0.56    &     1.10\\ 
   ESO572 &  2.10 &  1.87     &  0        &      -1.96$\dag$ &  0.48    &     14.81\\ 
  NGC7793 &  2.38 &    --     &  --         &      -1.34 &  0.04    &     0.06\\ 
  UGCA410 &  0.97 &  0.37     & 23          &   -1.84 &  0.58    &     36.52\\ 
  UGC9560 &  0.69 &  0.34     & 14          &   -2.02 &  0.51    &     36.63\\ 
  NGC1510 &  1.19 &  0.62     &  22       &     -1.71 &  0.43    &     8.42\\ 
  NGC1705 &  0.43 &  0.0      &  31         &      -2.42 &  0.81    &     4.20\\ 
  NGC4194 &  2.91 &  2.73     &  8          &   -0.26 &  0.02    &     10.14\\ 
   IC1586 &  2.02 &  1.37     & 25        &     -0.91 &  0.17    &     9.51\\ 
    MRK66 &  0.00 &  0.00     & 23          &      -1.94 &  0.42    &      13.50\\ 
   Haro15 &  0.73 &  0.27     &  18     &       -1.48 &  0.35    &     --\\ 
  NGC1140 &  0.93 &  0.62     &  13       &     -1.78 &  0.34    &     11.19\\ 
  NGC5253 &  0.32 &  0.12     &   9      &      -1.33 &  0.11    &     16.07\\ 
   MRK542 &  1.66 &     --    &  --         &      -1.32 &  0.32    &      11.23\\ 
  NGC6217 &  2.30 &  1.46     & 31          &   -0.74 &  0.05    &       1.30\\ 
  NGC7714 &  1.65 &  1.57     &  4          &   -1.23 &  0.09    &       18.40 \\ 
  NGC1614 &  3.90 &  3.47     & 17          &   -0.76 &  0.01    &     5.82    \\ 
  NGC6052 &  1.33 &  0.98     & 14          &   -0.72 &  0.06    &      8.68 \\ 
  NGC5860 &  3.77 &   --      &   --         &     -0.91 &  0.12    &         4.02  \\     
  NGC6090 &  1.80 &  1.62     &  8       &      -0.78$\ddag$ &  0.06    &     25.00\\ 
    IC214 &  2.54 &  2.08     & 18       &      -0.61 &  0.05    &      3.14\\ 
   MRK309 &  2.36 &   --      &  --         &       2.08$\dag$ &  0.03    &       3.68\\ 
  NGC3049 &  1.21 &    1.01   &  0        &      -1.14 &  0.13    &      4.49\\ 
  NGC4385 &  2.20 &  2.10     &  4          &      -1.02 &  0.09    &      6.60\\ 
  NGC5236 &  1.44 &  0.56     &  32      &      -0.83 &  0.06    &     0.44\\ 
  NGC7552 &  2.66 &  1.85     &  30      &       0.48 &  0.01    &   2.38  \\ \hline  
\end{tabular}  
\caption{\label{AvAv}   
Extinction properties, ultraviolet slope and photon escape probability  
of the galaxy sample.   
The second column is the extinction in magnitudes obtained   
with the MW extinction curve (see Appendix) without taking into account the effect of the   
underlying population. The third column is the  extinction in magnitudes   
obtained by combining the  MW extinction curve with the effect of an  
underlying population.  
The ratio between the equivalent widths of \Hb\ in absorption and    
in emission  
as defined in Section~\ref{under} is given in the fourth column.   
Values of the ultraviolet slope $\beta$ (see appendix) 
are from Meurer \etal\   \protect\shortcite{1999Meurer} except for   
those marked with a \ddag\ where the values were extracted from Calzetti et  
al. \protect\shortcite{1994Calzetti} and those marked   
with a \dag\ where we calculated ourselves the parameter $\beta$  
from the IUE data base  spectra, 
following Calzetti \etal\  \protect\shortcite{1994Calzetti} prescription.  
The escape probability $\epsilon$ is given in the sixth column (see text). 
Equivalent widths were calculated using Equation~\protect\ref{EW}   
except when marked with a   
$\star$ where they were taken from published spectroscopy.}  
\end{center}  
\end{table}  

Figure~\ref{F:Babs} shows the result of taking into account 
the corrections for Balmer absorptions due to an 
underlying stellar population. 
The medians of both  $\Delta OII$ and $\Delta H\alpha$ 
are close to zero indicating that including the underlying absorption 
correction brings into agreement the SFR in the optical with 
those in the FIR. 
At the same time $\Delta UV$ shows still a positive value indicating 
an excess with respect to the FIR estimate. 
We must remember that while the ratio of emission line fluxes to FIR flux 
is not very sensitive to changes in the photon escape from object to object, 
this is not the case for the ratio of UV continuum to FIR fluxes. 
The reason being that while  in the UV continuum we are 
detecting directly the escaped photons, i.e. those that do not 
heat the dust or ionize the gas, the emission lines and FIR fluxes 
are reprocessed radiation, i.e. the product of the radiation that does not 
escape the region. 
 
A striking aspect is the large reduction in the r.m.s. 
scatter in the  $\Delta UV$ from 0.70 before corrections to 
0.39 after corrections, i.e. about half the original value. 
This simple fact 
suggests the  goodness of the corrections  applied to the data.
This aspect is also illustrated  in Figure~\ref{SFRvsSFRC}
when compared to Figure~\ref{SFRvsSFR}.

\begin{figure*} 
\setlength{\unitlength}{1cm}    
\vspace{10. cm} 
\includegraphics{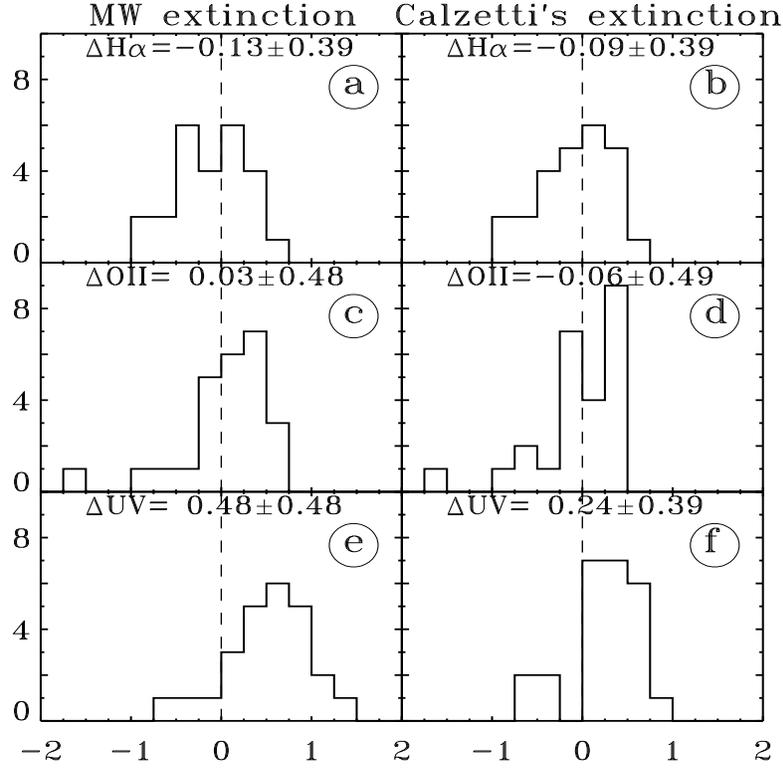} 
\caption{\label{F:Babs} The figure shows the histograms of  
the normalized SFR after the underlying stellar absorption effect is 
deducted from the extinction estimates, 
i..e. $A^*_V$ is used instead of $A_V$ (Section~\ref{under}). 
Panels a, c, e show the distributions after correcting 
the \Ha,  \OII\ and UV continuum  
using the MW extinction curve.  
b, d, and f show the result of using 
Calzetti's  extinction curve. 
The median and standard deviation are given for each case. The 
total number of objects is 25.}  
 
\end{figure*} 

\subsection{Photon Escape} 
\label{PhoSca}

As discussed above, while  $\Delta $OII and $\Delta $\Ha\ 
are not very sensitive to the fraction of escaped 
photons, $\Delta $UV is. 
The escape of UV photons can  
be quantified by the parameter $\epsilon $ \cite{1993Rowan} 
by assuming that

\begin{equation}\large  
\begin{array}{ll}  
L_{FIR} = \epsilon  F_{Bol} \\ 
L_{UV} = ( 1 - \epsilon )  F_{Bol} 
 
\end{array}  
\end{equation}  
where $( 1 - {\epsilon})$  represents the escaped fraction. 
 
Thus, the reprocessed fraction $\epsilon$ can be estimated 
through the observed 60$\mu$m and ultraviolet  
luminosities as 
 
\begin{equation}  
\epsilon = \frac{1}{1+\frac{\displaystyle L_{UV}}{\displaystyle L_{60\mu m}}}   
\end{equation}  
where the luminosities are given by  
\begin{equation}\large  
\begin{array}{ll}  
L_{UV}(erg\ s^{-1})&= F_\lambda(UV) \times 2000 \AA \\  
L_{60\mu m}(erg\ s^{-1})&= F_\nu(60\mu m)~  5 ~  
10^{12}Hz ~ 10^{-23}    
\end{array}  
\end{equation}  
 
The values of the UV flux in units of   
${\rm erg \: s}^ {-1} {\rm cm}^{-2} \AA ^{-1}$ and the fluxes at 
60$\mu$m in Jy     
 are given in Table~\ref{DATA} and the calculated values of the escape 
fraction $1 - \epsilon$, in Table~\ref{AvAv}.

The results of including the photon escape are also shown in the $\Delta $UV
histograms of Figure~\ref{F:GloRes}.
 As discussed above, under the simple assumption that 
dust and ionized gas have the same spatial distribution, the values of 
$\Delta$\Ha\ and $\Delta$OII do not change with respect 
to  Figure~\ref{F:Babs}.

In the left panel, the UV continuum was corrected    
using Steidel et~al. \shortcite{1999Steidel} simple approach 
of multiplying the observed UV flux by a fixed amount ($\times$5)  
which corresponds to the average correction found  
in a sample of local starburst galaxies  \cite{1994Calzetti} .  
The fact that the average correction for Calzetti's sample
and the average correction for our sample are almost identical  
suggests that both samples have been drawn from the same family
of objects, i.e.~they are similar samples.

The central panel shows the result of using the MW extinction law   
and applying to the observed values the expression~\ref{FtoF}  
where the visual extinction is Av$^\ast$ .
The panel at the right shows the result of
using the  extinction   
law given by  Calzetti \etal\  \shortcite{1994Calzetti}.  
Also in this case the visual extinction is Av$^\ast$.  

Inspection of the $\Delta$UV distribution in figures~\ref{F:NoCor},
\ref{F:Babs} and \ref{F:GloRes} 
shows that at least for the galaxies in the ``reference''  sample, 
the escape of photons is a minor effect compared to the underlying Balmer
absorption.

\begin{figure*} 
\setlength{\unitlength}{1cm}    
\vspace{8. cm} 
 
\includegraphics{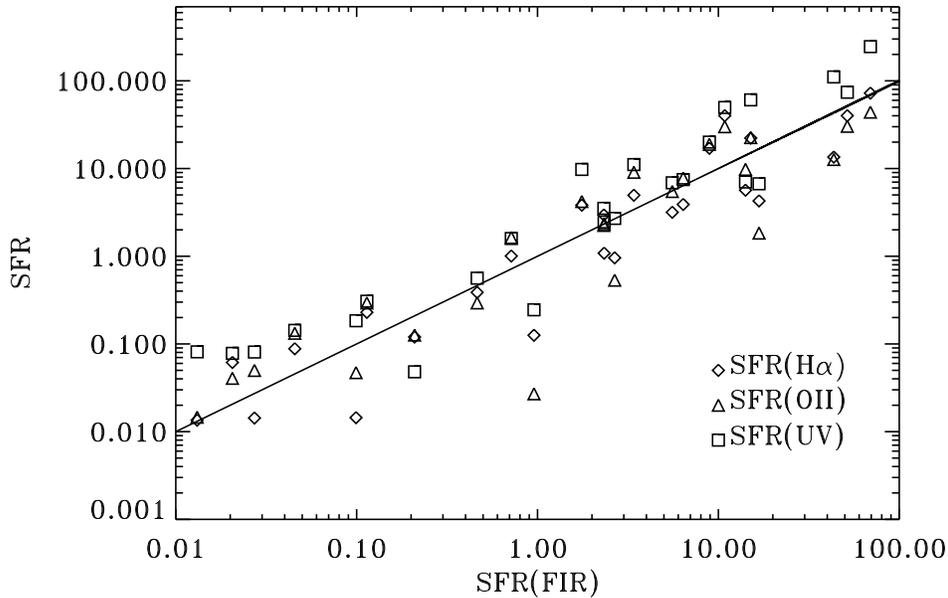}

\caption{\label{SFRvsSFRC} Corrected SFR estimators vs.~SFR(FIR).
The corrections include the underlying Balmer absorption (see 
Section~\ref{under}) and photon escape (see Section~\ref{PhoSca}. The 
solid line represents equal values.} 
\end{figure*} 
 

Comparing the MW and Calzetti's extinction corrections we can conclude 
that the corrections using the Calzetti's extinction curve give a 
SFR(UV) identical inside the errors  
to the SFR(FIR) plus a substantial reduction in the dispersion of the 
SFR estimates. 
These two points justify the use of Calzetti's extinction corrections 
in samples of starburst galaxies similar to the ones used in this work. 
It is a remarkable result that the application of our method 
succeeds in cutting down the scatter present in the original data 
to less than a half. These two results, i.e. the very close 
agreement between the SFR(UV) and SFR(FIR) 
plus the rather small scatter in the ratio of these two estimators 
gives confidence in the use of our approach to estimate dust extinction
and star formation rates in starburst galaxies.

\begin{figure*} 
\setlength{\unitlength}{1cm}    
\vspace{5.5 cm} 
\includegraphics{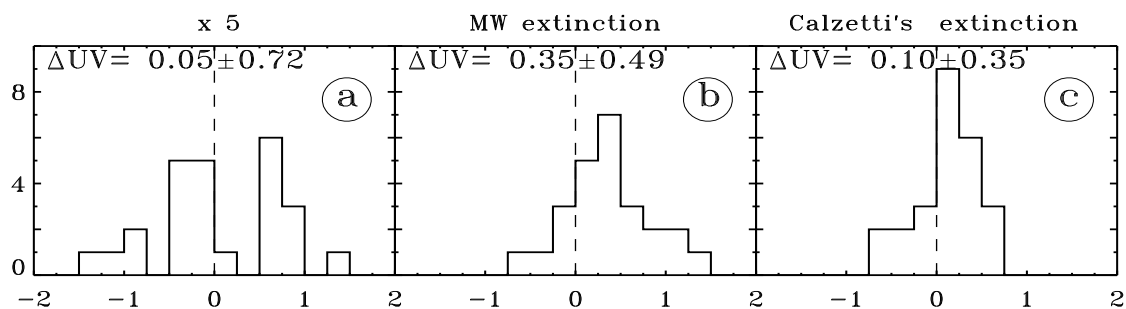} 
\caption{\label{F:GloRes} Histograms of  
the SFR given by the UV continuum
normalized to  the SFR(FIR).  
In the left panel we show the original data with the fixed correction 
given by Steidel et al. (1999).
The central  and right panels show the $\Delta$UV corrected for dust extinction   
computed including the underlying Balmer absorption (see Section~\ref{under})  
and photon escape. 
The median  and standard deviation are given for each case. 
The total number of objects is 25.}  
\end{figure*} 
  
\section{Unbiased SFR estimators } 
  
\vspace{1.5 cm}

The central result we have found is that the extinction correction including 
the effects of an underlying stellar Balmer absorption brings into agreement all four 
SFR estimators, and that the photon escape correction seems to play a minor role. 
 
We have seen that the SFR given  
by the UV continuum is equal to the SFR given by the FIR if the   
extinction correction is estimated using the  Calzetti extinction curve 
and including the underlying Balmer absorption corrections.   
It is reassuring that the simple inclusion of the underlying absorption correction 
brings into agreement theory and observations. 
We have used the results from the previous sections to obtain 
SFR estimators that are free from these systematic effects,

\begin{equation}\label{NewSFR} 
\left. 
\begin{array}{lcl} 
SFR(H\alpha )({\Myeareq})& =  1.1\times 10^{-41} & L_{H\alpha}{(\ergseq)} \\ 
SFR([O{\mbox{\footnotesize II}}] )({\Myeareq}) & =  8.4\times 10^{-41} & L_{OII}{(\ergseq)} \\  
SFR(UV)({\Myeareq}) & = 6.4 \times 10^{-28} & L_\nu({\ergseq Hz}^{-1}) \\   
\end{array}\right\} 
\end{equation} 
where the SFR(\Ha) is obtained from the reddening corrected \Ha\ luminosity while 
the expressions for the SFR(OII) and the SFR(UV) are for the observed fluxes. 
 
These estimators, (the SFR(OII) and the SFR(UV) from equation~\ref{NewSFR}) 
should be applied to samples where it 
is not possible to determine the extinction. 
If the objects have similar properties to our 
selected sample, then any systematic difference between the different estimators 
should be small. 
 
We have applied this new set of calibrations to  the SFR values given by  
different authors at different redshifts:  
Gallego \etal(1995) \Ha\ survey for the local Universe,  
Cowie \etal\shortcite{1995Cowie} \OII\ sample and Lilly \etal\  (1995)  
UV continuum  one between  redshifts of 0.2 and 1.5 and Connolly \etal\    
\protect\shortcite{1997Connolly} between 0.5 and 2;  
and  the UV points by Madau\etal\protect\shortcite{1996Madau}   
and Steidel\etal\protect\shortcite{1999Steidel} at redshift higher than 2.5.  
We give in Table~\ref{TablePapers} the complete list of surveys
of star formation at different redshifts that we have used plus their different tracers
and computed SFR.

The results are  plotted in Figure~\ref{MADAU}.  We 
have also plotted the  
results obtained by Hughes et al. (1998) based on SCUBA observations 
of the Hubble Deep Field, those  by Chapman et al. (2001)  
based on sub-millimeter observations of bright radio sources 
and those by Rowan-Robinson et al.~(1997) 
based on observations at 60 $\mu$m of the Hubble Deep Field. 
Clearly the mm/sub-mm results and  our ``unbiased'' 
results agree within the errors. 
The ``unbiased'' history of star formation is characterized  
by a large increase of the SFR density 
from z$\sim$0 to z$\sim$1 (a factor of about 20) and a slow 
decay from z$\sim$2 to z$\sim$5. 
 
The final corrected values are similar to other published results 
(e.g.~Somerville, Primack and Faber
2001). But the fact that our procedure removes systematic (or zero point) 
differences between
the different estimators, implies that the shape of the curve,  and therefore
the slopes between $z=0\, {\rm and}\, 1$ and $z=1\, {\rm and}\, 4$ are now 
better determined.

\begin{figure} 
\setlength{\unitlength}{1cm}    
\vspace{7. cm} 
\includegraphics{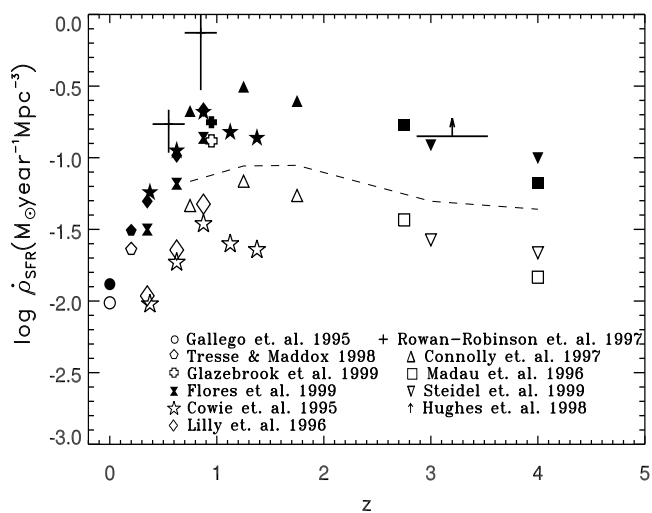} 
 \caption{\label{MADAU}   
The SFR density as a function of redshift.   
The solid and open symbols represent values corrected and uncorrected for reddening respectively, except
for the \Ha\ values (the two lowest z points) which are all reddening corrected.
The SFR in all cases are estimated  using equations~\ref{eq_Ha}, \ref{eq_OII}, \ref{eq_UV} and 
\ref{SFRIR} for open symbols and equations~\ref{NewSFR} for the filled symbols.   
The open circle is the SFR based on H$\alpha$   
emission line corrected by extinction from  Gallego\etal(1995) sample.    
The open stars are the SFR determined by  \OII\   
from the sample of Cowie\etal\protect\shortcite{1995Cowie}  
without any correction.  The
UV continuum samples are from   
Connolly \etal\  \protect\shortcite{1997Connolly} (triangles)  
Madau\etal\protect\shortcite{1996Madau} (squares)  
and Steidel\etal\protect\shortcite{1999Steidel} (upside down triangles).  
The upward arrow is the lower limit  
given by Hughes \etal\  \protect\shortcite{1998Hughes} based on sub-mm  
observations of the Hubble Deep Field. 
The crosses at z = 0.55 and 0.85 correspond to 
the SFR estimates by Rowan-Robinson \etal\ (1997).    
The dashed line is the 
lower limit given by Chapman \etal\ (2001)}  
\end{figure} 

\section{The history of star formation of the universe}

Let's now analyze with some detail the behaviour of the
dust corrected curve of Figure~\ref{MADAU}.

Figure~\ref{ourSFR} shows the corrected curve together with some 
representative theoretical models; (the details of the different
sets of data are listed in table~\ref{TablePapers}). 
Our  basic assumption is that the main results from our sample of 
galaxies in the nearby universe are also applicable to 
samples of starforming galaxies at intermediate and high redshift.

An important result is that within observational errors, our calibrations
succeed in bringing into reasonable agreement the FIR determinations
with the optical/UV ones. 
A first implication of this agreement is that 
large dust extinction corrections are not favoured.
It can be seen that while the general trend present in previous
determinations of the star formation history is preserved, the
rise from $z=0$ to $z=1$ is steeper than in most previous work.
The procedure we have used minimizes any systematic differences between
the different redshift ranges and implies that, provided the composition
of the sample of objects at intermediate and high redshift is similar 
to that of our nearby sample, the shape of the curve would be preserved.

The general behaviour of the star formation history of the universe
can be represented with 
a steep rise in the SFR rate
from $z=0$ to $z=1$ with SFR $\propto (1+z)^{4.5}$
and a decline from $z=1$ to  $z=4$ where  SFR $\propto (1+z)^{-1.5}$ (see solid lines
in  Figure~\ref{ourSFR}). The steep increase to $z=1$ is similar to the 
slope found by Glazebrook et al. (1999) using the 
\Ha\ luminosities only.

We are confident that the  procedure presented in this paper guarantees 
the removal of systematic
differences between the star formation estimators when they are applied to
samples qualitatively similar to ours. This in turn means that the general
shape and in particular the slope from $z = 0$ to $z = 1$ and from 
$z = 1$ to $z = 4$ are well determined, although the exact value of the
$z = 0$ to $z = 1$ slope is still very dependent on the exact value
of the SFR in the nearby universe.

We also included in Figure~\ref{ourSFR} the results from some representative 
semi-analytical models of galaxy formation corresponding to the
hierarchical cold dark matter scenario. The dotted lines correspond to 
Somerville et al. (2001) $\tau$CDM models with two different normalizations 
while the dashed line corresponds to  model C from Baugh et al. (1998).

Whatever the local value of the SFR is, the comparison with these
models of galaxy formation shows a general agreement in the shape
plus what we can call a local excess in the models, i.e.~the models have a
shallower slope between $z = 0$ to $z = 1$  than the data suggest
or equivalently, that present semi-analytical models of galaxy formation 
seem unable to reproduce the sharp increase in the SFR from $z=0$ to $z=1$.

\begin{table*}
\begin{center}
\begin{tabular}{|c|c|c|c|c|c|l|}\hline
Redshift   & SFR     &  log ($\cal L$)              & log(SFR)    & log(SFR)
&\footnotesize Log Error&  \ \ \ \ \ \ \ \ \ \ References \\
Range     & Tracer  &                       & (Standard)  & (Unbiased)   &
\ & \\ \hline
0.0-0.045 & \Ha     &  39.09  &  -2.01      & -1.87       & 0.2
&\protect\cite{1995Gallego}\\ 
0.1-0.3   & \Ha     &  39.47  &  -1.63      & -1.50       & 0.04
&\protect\cite{1998Tresse}\\
0.75-1.0  & \Ha     &  40.22  &  -0.75      & -0.88       & 0.17
&\protect\cite{1999Glazebrook}\\
0.25-0.50 & FIR     &  41.85  &  -1.50      & -1.50       & 0.26 &
\protect\cite{1999Flores}\\
0.50-0.75 & FIR     &  42.17  &  -1.18      & -1.18       & 0.26 &
\protect\cite{1999Flores}\\
0.75-1.00 & FIR     &  42.49  &  -0.86      & -0.86       & 0.26 &
\protect\cite{1999Flores}\\
0.25-0.50 & OII     &  38.83  &  -2.02      & -1.24       & 0.1  &
\protect\cite{1995Cowie}\\
0.50-0.75 & OII     &  39.12  &  -1.73      & -0.95       & 0.1  &
\protect\cite{1995Cowie}\\
0.75-1.00 & OII     &  39.39  &  -1.46      & -0.68       & 0.1  &
\protect\cite{1995Cowie}\\
1-1.25    & OII     &  39.25  &  -1.60      & -0.82       & \dag &
\protect\cite{1995Cowie}\\
1.25-1.50 & OII     &  39.21  &  -1.64      & -0.86       & \dag &
\protect\cite{1995Cowie}\\
0.25-0.50 & UV      &  25.89  &  -1.96      & -1.30       & 0.07 &
\protect\cite{1996Lilly}\\
0.50-0.75 & UV      &  26.21  &  -1.64      & -0.98       & 0.08 &
\protect\cite{1996Lilly}\\
0.75-1.00 & UV      &  26.53  &  -1.32      & -0.66       & 0.15 &
\protect\cite{1996Lilly}\\
0.4-0.7   & FIR     &  42.64  &  -0.76      & -0.76       & (+0.1)(-0.2) &
RR97\\       
0.7-1.0   & FIR     &  43.23  &  -0.13      &  -0.13      & (+0.4)(-0.2) &
RR97\\
0.5-1.0   & UV      &  26.52  &  -1.33      &  -0.63      & 0.15 &
\protect\cite{1997Connolly}\\
1.0-1.5   & UV      &  26.69  &  -1.16      &  -0.50      & 0.15 &
\protect\cite{1997Connolly}\\ 
1.5-2.0   & UV      &  26.59  &  -1.26      &  -0.60      & 0.15 &
\protect\cite{1997Connolly}\\
2.0-4.0   & FIR     &  42.80  &  -0.85      &  -0.85      & \dag &
\protect\cite{1998Hughes}\\
2.0-3.5   & UV      &  26.42  &  -1.43      &  -0.77      & 0.15 &
\protect\cite{1996Madau}\\
3.5-4.5   & UV      &  26.02  &  -1.83      &  -1.17      & 0.2  &
\protect\cite{1996Madau}\\
2.8-3.3   & UV      &  26.28  &  -1.57      &  -0.91      & 0.07 &
\protect\cite{1999Steidel}\\
3.9-4.5   & UV      &  26.19  &  -1.66      &  -1.00      & 0.1  &
\protect\cite{1999Steidel}\\ \hline
\end{tabular}
\caption[SFR for surveys at different redshifts.]
{\label{TablePapers} SFR for surveys at different redshifts.
The mean luminosity density ($\cal L$) is given in units of \ergs Mpc$^{-3}$
for the case of the \Ha\ and OII\  emission lines and the FIR.
The UV is given in \ergs Hz$^{-1}$Mpc$^{-3}$. 
The SFR referred to as standard are given by the expressions by 
Kennicutt (1998) and described in Section~\ref{methods}. The 
SFR referred to as unbiased are the ones obtained by using
Equation~\ref{NewSFR}. 
In both cases, the SFR are given in units of $\Myeareq$Mpc$^{-3}$.}
\end{center}
\end{table*}

\begin{figure} 
\setlength{\unitlength}{1cm}    
\vspace{7. cm} 
\includegraphics{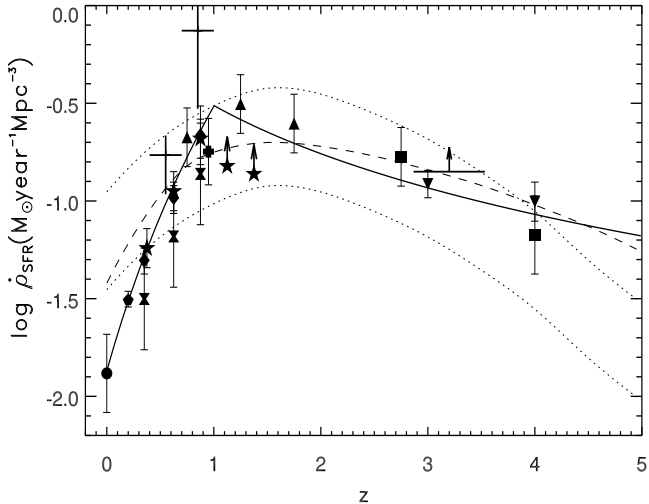} 
 \caption{\label{ourSFR}   
The unbiased SFR density as a function of redshift. The dotted lines represent
the two normalizations in Somerville et al. (2001) $\tau$CDM models, and the 
dashed line is model C from Baugh et al. (1998). The solid lines represent
SFR $\propto (1+z)^{4.5}$ from $z=0$ to $z=1$ and SFR $\propto (1+z)^{-1.5}$
for $z>1$ .}
\end{figure} 

\section{Conclusions}  
  
One aspect that has generated many discussions regarding the SF history 
diagram in Cosmology, is the lack of confidence on the reddening corrections. 
This is compounded with the low level of agreement found until now
between the optical/UV and FIR determinations of the SFR. 
 
In the first part of this paper, we have investigated the possible systematic 
differences between 
SFR estimators by applying them to a sample of nearby 
star forming galaxies with good photometric data 
from the UV to the FIR. 

We found that the main source of systematic differences among the
SFR rate estimators is related to the presence of stellar  Balmer absorption
in the spectrum of emission line galaxies. The main effect of the Balmer absorptions
is to produce an overestimate of the reddennnig when their effect
is not included.
We showed that taking into account the underlying Balmer absorptions
effect in the estimates of reddening,  removes most of the systematic
differences between the SFR estimators in the
optical/UV and FIR.  
Furthermore the scatter of the SFR estimations
is considerably reduced by the application of the corrections.
We also found that the escape of photons plays a minor role  compared
to that of the Balmer absorptions.
These results give renewed confidence to the estimates of SFR 
for star forming galaxies in general and for samples 
similar to the one presented here in particular.

{\bf Thus, our central result is that the extinction correction including 
the effects of an underlying stellar Balmer absorption brings into agreement all four 
SFR estimators, and that the photon escape correction seems to play a minor role.}

In the second part of the paper
we used the average results for our sample to construct a set 
of ``unbiased'' SFR estimators. 
These ``unbiased'' SFR estimators expressions include statistically 
the underlying Balmer absorption and 
photon escape corrections to the extinction estimates and bring 
the four SFR estimators studied here into the same system. 
We thus obtained
consistent results between the SFR estimators in the 
optical/UV and FIR.  
The application of these ``unbiased'' SFR estimators to a compilation
of surveys has produced a SFR history of the universe where all surveys results
agree whitin the errors. Particularly important is the level of agreement 
achieved between the FIR/mm and optical/UV SFR results.

Our ``new" and unbiased  SFR history of the universe shows 
a steep rise in the SFR rate
from $z=0$ to $z=1$ with SFR $\propto (1+z)^{4.5}$
followed by a mild decline for $z>2$ where  SFR $\propto (1+z)^{-1.5}$.
The steep increase to $z=1$ seems in line with recent determinations of the
SFR using only the \Ha\ estimator.
Most galaxy formation  models tend to have a much flatter slope 
from $z=0$ to $z=1$.

\section{Acknowledgments}  
We acknowledge fruitful conversations with 
Miguel Mas-Hesse,  
Daniela Calzetti, Piero Madau, Max   
Pettini, Divakara Mayya, Enrique P\'erez, Rosa Gonz\'alez-Delgado, 
 David Hughes and Enrique Gazta\~naga and useful suggestions from an 
anonymous referee.  
Daniel Rosa Gonz\'alez gratefully acknowledges   
a research grant from the INAOE Astrophysics Department, a studentship from  
CONACYT, the Mexican Research Council, as part of ET research grant  
\#~32186-E, and an EC Marie Curie studentship at the IoA Cambridge.

\newpage 
 
\appendix 
\section{Dust Extinction Corrections to the Observed Fluxes}   
\label{SEC:CORR}

Two different extinction curves are used: the Milky Way   
extinction law  (MW) given by Seaton \shortcite{1979Seaton}  
and Howarth \shortcite{1983Howarth} and the Large Magellanic   
Cloud one (LMC)  
given by Howarth \shortcite{1983Howarth}. The main   
parameters of both laws are given in Table~\ref{T:EXT}.  
  
\subsection{Dust Extinction Corrections to the  
Continuum  Fluxes}   
\label{SUBSEC:dustCORR}  
  
Calzetti and collaborators developed an empirical method to  
estimate the UV extinction (Calzetti, Kinney and Storchi-Bergmann, 
 1994). They found that  
the power-law index $\beta$ in the ultraviolet defined  
as $F_\lambda\propto \lambda^\beta$      
is well correlated with the difference in  optical depth between  
\Ha\ and \Hb\ defined as  
 $\tau_B^l={\rm ln}\left( \frac{F(H\alpha)/F(H\beta)}{2.86}\right)$  
where $F(H\alpha)$ and $F(H\beta)$ are the intensities of the  
\Ha\ and \Hb\ emission lines respectively.  
This  correlation, which is linear and independent of the adopted   
extinction law is given by,   
\begin{equation}  
\label{BTAU}  
\beta = (1.76 \pm 0.25)\tau_B^l - (1.71 \pm 0.12)  
\end{equation}  
The parameter $\beta$ is obtained by fitting the power law to  
the IUE ultraviolet spectra. Calzetti \etal\    
(1994) and Meurer, Heckman and Calzetti (1999) 
values of $\beta$ are presented   
in Table \ref{AvAv} as well as our estimate for CAM0840,   
CAM1543, TOL1247, ESO572 and MRK309.   
  
The effect of reddening using different   
dust spatial distributions can be estimated from Equation~\ref{BTAU}  
by comparing ultraviolet with optical   
spectra. Calzetti \etal\    
\shortcite{1994Calzetti}   
estimate the optical depth $\tau_\lambda$ by solving the   
transfer equation for five different geometries, uniform or clumpy   
dust   
screen, uniform or clumpy scattering slab and internal dust.  
The  uniform dust screen constitutes the easiest case where the optical   
depth is related to the visual extinction by    
$$\tau_\lambda = 0.921 k(\lambda) E(B-V)$$  
where E(B-V) is the  colour excess and $k(\lambda)$ is the extinction  
law. For the other geometries, apart from the assumed   
extinction law, the optical depth is a function of dust   
parameters such as the albedo, the  
phase parameter or the number of clumps.   
After comparing synthetic extinction corrected spectra  
with observations of emission line galaxies
Calzetti \etal\  \shortcite{1994Calzetti}   
conclude that none of the adopted geometries   
combined with the {\it standard} MW and LMC extinction laws    
could explain  
the observed  tight relation between $\tau_B^l$ and $\beta$ and  
proposed an empirical extinction law obtained from  
IUE spectra of a sample of nearby  starburst galaxies.   
  
Calzetti \etal\  \shortcite{1994Calzetti} created 6 different   
templates averaging galaxies  
with the same amount of dust (judging by their Balmer decrements).  
The template with  $\tau_B^l=0.05$ is taken as the reference  
one (free of dust).   
An optical depth, $\tau_n(\lambda)$, is calculated for each template by   
comparing the observed  
fluxes, $F_n(\lambda)$ and $F_1(\lambda)$,  
\begin{equation}  
\tau_n(\lambda)= - \ln \frac{F_n(\lambda)}{F_1(\lambda)}  
\end{equation}  
where the subindex 1 corresponds to the dusty free template and   
the subindex n corresponds to the n template.  
For each template a rescaled optical depth  can be defined    
\begin{equation}  
Q_n(\lambda)= \frac{\tau_n(\lambda)}{\tau^l_{Bn}-\tau^l_{B1}(\lambda)}  
\end{equation}  
Averaging this quantity, Calzetti et al. (1994) found an extinction  
curve, $Q(\lambda)$   
which can be transformed to $k(\lambda)$ (e.g.  
 Seaton, 1979) by,  
\begin{equation}  
Q(\lambda)= \frac{k(\lambda)}{k(H\beta)-k(H\alpha)}  
\end{equation}  
where the difference $k(H\beta)-k(H\alpha$) is given by the  
Seaton (1979) extinction curve.  
The observed ultraviolet flux is related to  the emitted one  
by, 
\begin{equation}\label{FtoF}   
F_{obs}(\lambda) = F_o(\lambda) 10^{-0.4 A{\rm v}\: k (\lambda)/R}   
\end{equation}  
\noindent 
where   $ k (\lambda)$ is given by  \cite{1999Calzetti} 
\begin{equation}\label{k}  
k(\lambda) = -2.156 + 1.509/\lambda - 0.198/  
\lambda^2 +0.011/\lambda^3   
\end{equation}  
valid  for the range 0.12 $\mu$m $<  \lambda <$ 0.63 $\mu$m.  
  
The obtained extinction curve can be considered as  
an average of the different dust distributions described by    
Calzetti\etal\shortcite{1994Calzetti}.   
  
In order to correct the ultraviolet flux using this procedure it is   
necessary  
to estimate Av from  the observed  
\Ha $/$\Hb\  ratios and then apply Equation \ref{k} to the observed  
ultraviolet fluxes.   
Physically this correction is understood assuming that   
the ionized gas is more affected by extinction than the  
stars which are producing the observed UV flux \cite{1994Calzetti}. 
 
No corrections were applied to the IR data.  
  
The different  
extinction curves are plotted in Figure \ref{F:EXT}.    
  
\begin{table*}  
\begin{center}  
\begin{tabular}{|l|c|c|c|c|c|}\hline  
\ &$ k$(\Ha)-$k$(\Hb)  & $k$(\Hg) - $k$(\Hb) &  $R$=$k$(5464 \AA) & $k$(\OII) & $k$(1700 \AA) \\   
MW       &  - 1.25        &   0.45           &   3.2        &  4.67    & 7.80 \\   
LMC      &  - 1.18        &  0.48            &   3.2        &  4.86    & 9.54 \\    
Calzetti &  - 0.58        &  0.23            &   2.7        &  3.46    & 5.10 \\ \hline  
\end{tabular}  
\caption{\label{T:EXT} Adopted values for the extinction curves.}  
\end{center}  
\end{table*}  
  
\begin{figure} 
\setlength{\unitlength}{1cm}    
\vspace{7.5 cm} 
\includegraphics{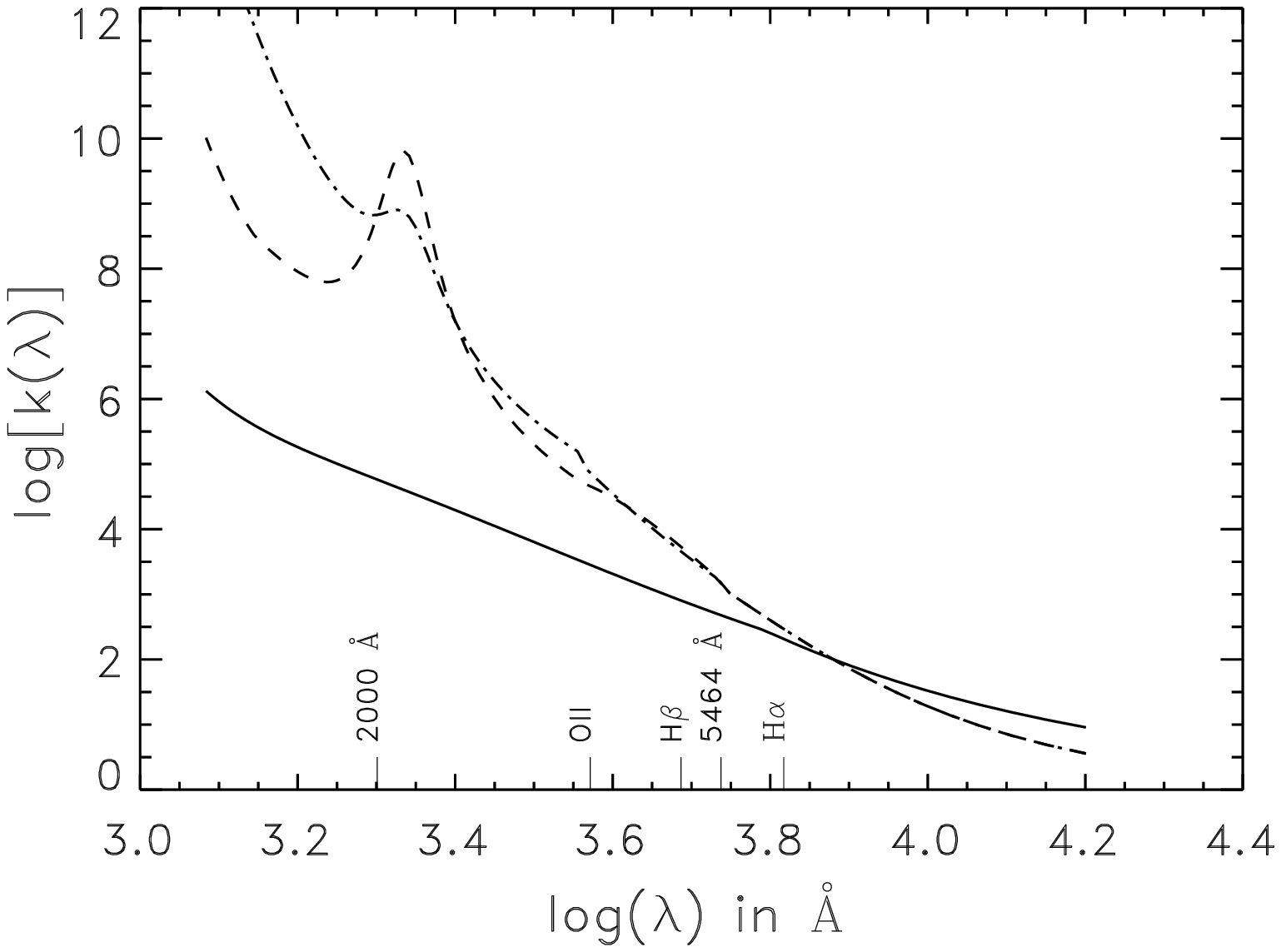} 
\caption{\label{F:EXT} Adopted extinction curves. The solid line is the empirical relation   
given by Calzetti \protect\shortcite{1999Calzetti}. The dashed line is the  
curve for the   
MW (Seaton \protect\shortcite{1979Seaton}  
and Howarth \protect\shortcite{1983Howarth})  
and the dot-dashed line is the curve for the LMC \protect\cite{1983Howarth}.}  
\end{figure} 

\subsection{Dust extinction corrections to the Emission Line Fluxes}  
\label{C_EL}  
  
Extinction affects the emission lines in different degrees depending  
on wavelength. Corrections are usually obtained from the observed  
ratio of Balmer lines, the intrinsic ratio,     
and an adopted interstellar extinction curve.  
  
The ratio between the intensity of a given line $F(\lambda)$    
and the intensity of \Hb, $F$(\Hb) can be expressed by:   
 \begin{equation}   
\label{ext}   
\frac{F(\lambda)}{F(H\beta)} = \frac{F_o(\lambda)}{F_o(H\beta)}   
 10 ^{-0.4 {\rm Av} [k(\lambda) - k(H\beta)]/R}   
\end{equation}   
where the difference $k(\lambda) - k(H\beta)$ is tabulated for different  
extinction laws (Table \ref{T:EXT}).  
The total visual extinction Av, depends on the observed object (see Table~\ref{AvAv}).   
The subindex $o$ indicates  the unreddened values.   
We use as reference the theoretical ratio for    
Case B recombination $F_o$(\Ha)/$F_o$(\Hb) =2.86 and   
$F_o$(\Hg)/$F_o$(\Hb)=0.47 \cite{Osterbrock}.   
The observed flux ratios can be expressed as a function of the   
theoretical ratios and the visual extinction,   
\begin{equation}  
\label{VEC_EQ}  
\begin{array}{ccc}  
{\rm log} \frac{F(H\alpha)}{F(H\beta)} = & {\rm log\ } 2.86\ - & 0.4 [k(H\alpha) -   
k(H\beta)]{\rm Av}/R \\  
{\rm log} \frac{F(H\gamma)}{F(H\beta)} = & {\rm log\ } 0.47\ - & 0.4 [k(H\gamma) -   
k(H\beta)]{\rm Av}/R  
\end{array}  
\end{equation}  
This Equation was  used to analyze the presence of   
an underlying stellar population in Section \ref{under}.

\bsp  
\label{lastpage}  
\end{document}